\documentclass[prd,preprint,tightenlines,floatfix,
showpacs,preprintnumbers,nofootinbib,eqsecnum,superscriptaddress]{revtex4}

\usepackage[T1]{fontenc}		

\usepackage{amsmath,amsfonts,amssymb,amstext,mathrsfs}
\usepackage{mathpazo}

\usepackage[dvips]{graphicx}
\usepackage{epsf,float}
\usepackage{revsymb}

\usepackage{dcolumn}
\usepackage{braket}
\usepackage{color,xcolor}
\usepackage{graphicx}
\usepackage{subfigure}
\usepackage{multirow}
\usepackage{tabularx}
\usepackage{pstricks}
\usepackage[section]{placeins}
\usepackage{booktabs}
\usepackage{array}

\usepackage{hyperref}

\newcommand{\Pom}{\mathbb{P}}

\newcommand{\Reg}{\mathbb{R}}

\begin{document}

\title{
Exclusive diffractive production of $\pi^{+}\pi^{-}\pi^{+}\pi^{-}$\\
via the intermediate $\sigma\sigma$ and $\rho\rho$ states
in proton-proton collisions \\
within tensor pomeron approach}

\author{Piotr Lebiedowicz}
 \email{Piotr.Lebiedowicz@ifj.edu.pl}
\affiliation{Institute of Nuclear Physics PAN, PL-31-342 Krak\'ow, Poland}

\author{Otto Nachtmann}
 \email{O.Nachtmann@thphys.uni-heidelberg.de}
\affiliation{Institut f\"ur Theoretische Physik, Universit\"at Heidelberg,
Philosophenweg 16, D-69120 Heidelberg, Germany}

\author{Antoni Szczurek
\footnote{also at University of Rzesz\'ow, PL-35-959 Rzesz\'ow, Poland.}}
\email{Antoni.Szczurek@ifj.edu.pl}
\affiliation{Institute of Nuclear Physics PAN, PL-31-342 Krak\'ow, Poland}

\begin{abstract}
We present first predictions of the cross sections and 
differential distributions for the central exclusive reaction
$pp \to pp \pi^{+} \pi^{-} \pi^{+} \pi^{-}$ being studied at RHIC and LHC.
The amplitudes for the processes are formulated in terms 
of the tensor pomeron and tensor $f_{2 \Reg}$ reggeon exchanges
with the vertices respecting the standard crossing 
and charge-conjugation relations of Quantum Field Theory.
The $\sigma \sigma$ and $\rho \rho$ contributions 
to the $\pi^{+} \pi^{-} \pi^{+} \pi^{-}$ final state
are considered focussing on their specificities.
The correct inclusion of the pomeron spin structure seems crucial for
the considered sequential mechanisms in particular for the $\rho\rho$ contribution
which is treated here for the first time.
The mechanism considered gives a significant contribution to the
$pp \to pp \pi^{+} \pi^{-} \pi^{+} \pi^{-}$ reaction.
We adjust parameters of our model to 
the CERN-ISR experimental data
and present several predictions for the STAR, ALICE, ATLAS and CMS experiments.
A measurable cross section of order of a few $\mu b$ is obtained 
including the experimental cuts relevant for the LHC experiments.
We show the influence of the experimental cuts 
on the integrated cross section and on various differential distributions.
\end{abstract}


\maketitle

\section{Introduction}

Last years there was a renewed interest in exclusive production 
of two meson pairs (mostly $\pi^+ \pi^-$ pairs) at high energies
related to planned experiments at RHIC \cite{Adamczyk:2014ofa}, 
Tevatron \cite{Aaltonen:2015uva,Albrow_Project_new}, and LHC
\cite{Schicker:2014aoa, Staszewski:2011bg, CMS:2015diy}.
From the experimental point of view the exclusive processes are important
in the context of resonance production, in particular, in searches for glueballs.
The experimental data on central exclusive $\pi^{+}\pi^{-}$ production
measured at the energies of the ISR, RHIC, Tevatron, and the LHC collider 
all show visible structures in the $\pi^{+}\pi^{-}$ invariant mass.
It is found that the pattern of these structures seems to depend on experiment.
But, as we advocated in Ref.~\cite{Lebiedowicz:2016ioh},
this dependence could be due to the cuts used in a particular experiment
(usually these cuts are different for different experiments).

So far theoretical studies concentrated on two-pion continuum production.
Some time ago two of us have formulated a Regge-type model with parameters 
fixed from phenomenological analysis of total and elastic $NN$ and $\pi N$ 
scattering \cite{Lebiedowicz:2009pj}.
The model was extended to include also absorption effects 
due to proton-proton interaction \cite{Lebiedowicz:2011nb,Lebiedowicz:2011tp}.
In Ref.~\cite{Lebiedowicz:2011nb} the exclusive reaction 
$pp \to pp \pi^{+}\pi^{-}$
constitutes an irreducible background to the scalar $\chi_{c0}$ meson 
production.
These model studies were extended also to $K^{+}K^{-}$ 
production \cite{Lebiedowicz:2011tp}.
For a related work on the exclusive reaction 
$pp \to nn \pi^{+}\pi^{+}$, see \cite{Lebiedowicz:2010yb}.
A revised view of the absorption effects including the 
$\pi N$ nonperturbative interactions
was presented very recently \cite{Lebiedowicz:2015eka}.
Such an approach gives correct order of magnitude cross sections,
however, does not include resonance contributions which interfere 
with the continuum. 

It was known for a long time that the frequently 
used vector-pomeron model has problems
considering a field theory point of view.
Taken literally it gives opposite signs for $pp$ and $\bar{p}p$ total 
cross sections.
A way out of these problems was already shown in \cite{Nachtmann:1991ua}
where the pomeron was described as a coherent superposition of exchanges
with spin 2 + 4 + 6 + ... . 
The same idea is realised 
in the tensor-pomeron model formulated in \cite{Ewerz:2013kda}.
In this model pomeron exchange can effectively be treated as the
exchange of a rank-2 symmetric tensor.
The corresponding couplings of the tensorial object 
to proton and pion were worked out.
In Ref.~\cite{Lebiedowicz:2013ika} the model was applied to 
the diffractive production of several scalar and pseudoscalar mesons 
in the reaction $p p \to p p M$.
In \cite{Bolz:2014mya} an extensive study of the photoproduction reaction
$\gamma p \to \pi^{+} \pi^{-} p$ in the framework
of the tensor-pomeron model was presented.
The resonant ($\rho^0 \to \pi^{+}\pi^{-}$) and non-resonant (Drell-S\"oding)
photon-pomeron/reggeon $\pi^{+} \pi^{-}$ production in $pp$ collisions
was studied in \cite{Lebiedowicz:2014bea}.
The exclusive diffractive production of $\pi^{+} \pi^{-}$ continuum 
together with the dominant scalar $f_{0}(500)$, $f_{0}(980)$, 
and tensor $f_{2}(1270)$ resonances was studied by us
very recently in Ref.~\cite{Lebiedowicz:2016ioh}.


The past program of central production of pairs of mesons was
concentrated on the discussion of mesonic resonances.
The low-energy program of studying meson excitations can be
repeated at the LHC, where we expect dominance of one 
production mechanism only, two-pomeron exchange.

The identification of glueballs can be very difficult.
The partial wave analyses of future experimental data of the STAR,
ALICE, ATLAS and CMS Collaborations could be used in this context.
Also the studies of different decay channels in central
exclusive production would be very valuable.
One of the possibilities is the $p p \to p p \pi^+ \pi^- \pi^+ \pi^-$
reaction being analysed by the ATLAS, CMS and ALICE Collaborations at the LHC.
Identification of the glueball-like states in this channel requires
calculation/estimation of the four-pion background from other sources.

Pairs of $\rho^{0} \rho^{0}$ (giving four pions) can be produced also in photon-hadron interactions
in a so-called double scattering mechanism.
In Ref.~\cite{Goncalves:2016ybl} double vector meson production in 
photon-photon and photon-hadron interactions in $pp$/$pA$/$AA$ collisions was studied.
In heavy ion collisions the double scattering mechanism is very important
\cite{KlusekGawenda:2013dka,Goncalves:2016ybl}.
In proton-proton collisions, for instance at the center-of-mass energy of $\sqrt{s} = 7$~TeV,
total cross sections for the double $\rho^{0}$ meson production,
taking into account the $\gamma \gamma$ and double scattering mechanisms,
were estimated \cite{Goncalves:2016ybl} to be of 182~$pb$ and 4~$pb$, respectively.

In the present paper we wish to concentrate on the four charged pion
continuum which is a background for future studies of diffractively produced resonances.
We shall present a first evaluation of the four-pion continuum 
in the framework of the tensor pomeron model 
consistent with general rules of Quantum Field Theory. 
Here we shall give explicit expressions 
for the amplitudes of $\rho \rho \equiv \rho(770) \rho(770)$ and 
$\sigma \sigma \equiv f_0(500) f_0(500)$
production with the $\rho$ and $\sigma$ decaying to $\pi^+ \pi^-$.
We shall discuss their specificity and relevance for
the $p p \to p p \pi^+ \pi^- \pi^+ \pi^-$ reaction.
In the Appendix~\ref{sec:diagram2} we present the formulas of
the double-pomeron exchange mechanism for the exclusive production of scalar resonances
decaying into $\sigma \sigma$ and/or $\rho \rho$ pairs.
\section{Exclusive diffractive production of four pions}
\label{sec:section_two_meson}

\begin{figure}
\includegraphics[width=6.cm]{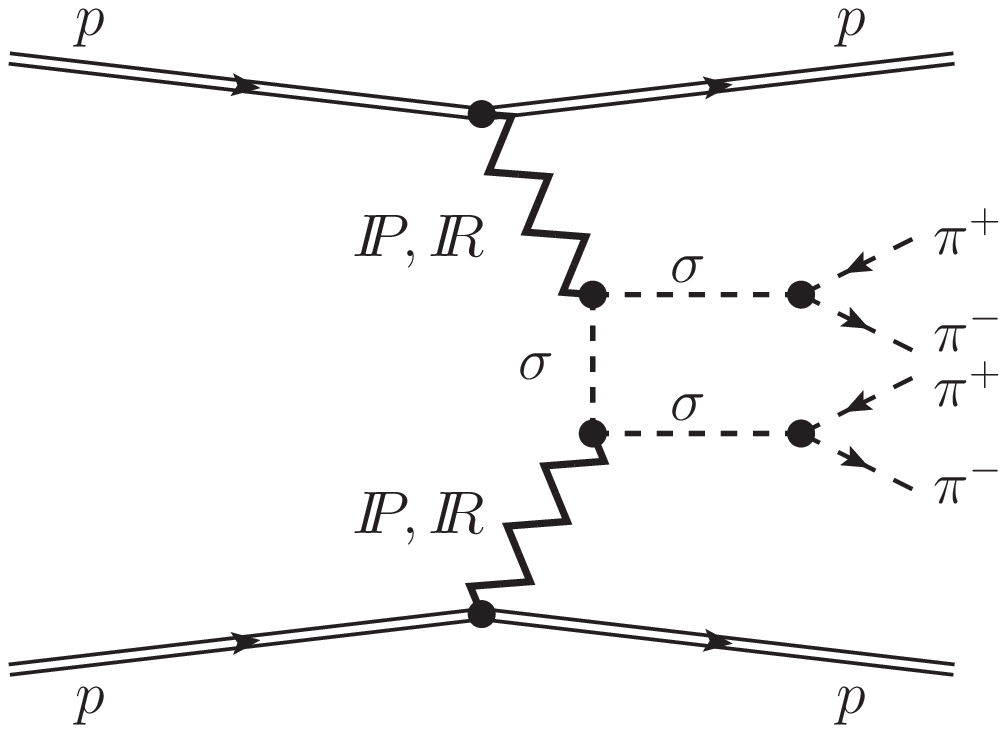} 
\includegraphics[width=6.cm]{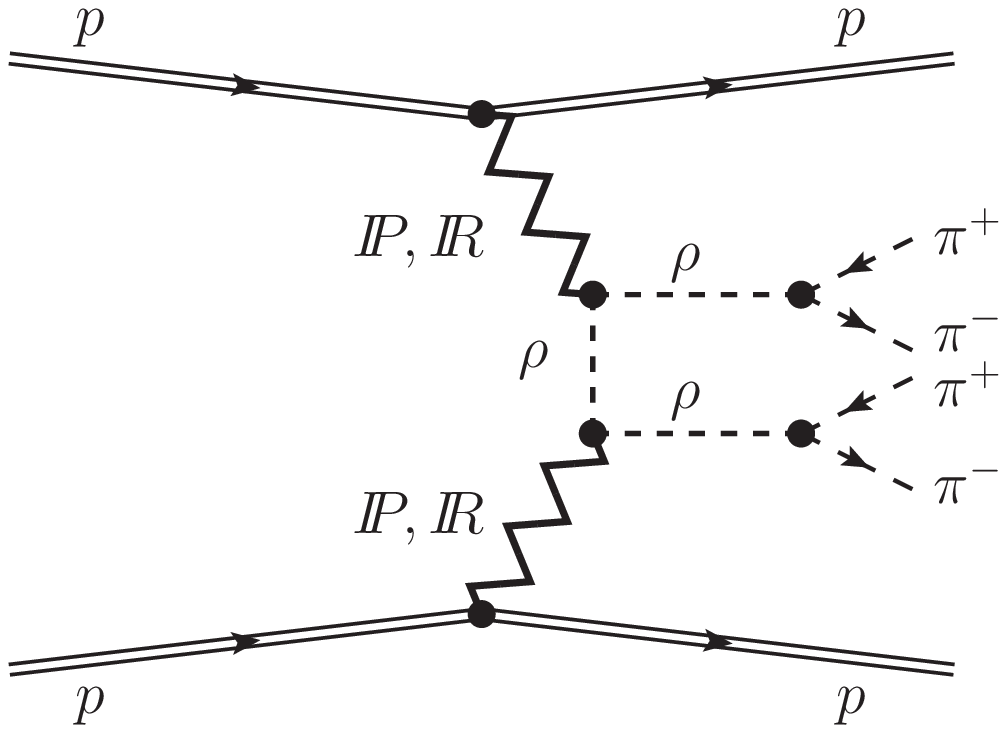} 
  \caption{\label{fig:4pi_diagram1}
  \small The ``Born level'' diagrams for double-pomeron/reggeon
  central exclusive $\sigma \sigma$ (left diagram) and $\rho \rho$ (right diagram) production 
  and their subsequent decays into $\pi^+ \pi^- \pi^+ \pi^-$ in proton-proton collisions.
}
\end{figure}

In the present paper we consider the $2 \to 6$ processes from the diagrams 
shown in Fig.~\ref{fig:4pi_diagram1},
\begin{equation} 
\begin{split}
& p  p \to p  p \,\sigma  \sigma \to p  p \, \pi^{+}  \pi^{-}  \pi^{+}  \pi^{-}\,, \\
&p  p \to p  p \, \rho  \rho \to p  p\,  \pi^{+}  \pi^{-}  \pi^{+}  \pi^{-}\,.
\label{2to6_reaction}
\end{split}
\end{equation} 
That is, we treat effectively the $2 \to 6$ processes (\ref{2to6_reaction})
as arising from $2 \to 4$ processes,
the central diffractive production of
two scalar $\sigma$ mesons and two vector $\rho$ mesons 
in proton-proton collisions.
To calculate the total cross section for the $2 \to 4$ reactions
one has to calculate the 8-dimensional phase-space integral numerically \cite{Lebiedowicz:2009pj}
\footnote{In the integration over four-body phase space
the transverse momenta of the produced particles ($p_{1t}$, $p_{2t}$, $p_{3t}$, $p_{4t}$),
the azimuthal angles of the outgoing protons ($\phi_{1}$, $\phi_{2}$)
and the rapidity of the produced mesons ($y_{3}$, $y_{4}$)
were chosen as integration variables over the phase space.}.
Some modifications of the $2 \to 4$ reaction are needed
to simulate the $2 \to 6$ reaction with 
$\pi^{+} \pi^{-} \pi^{+} \pi^{-}$ in the final state.
For example, one has to include in addition a smearing of the 
$\sigma$ and $\rho$ masses due to their instabilities.
Then, the general cross-section formula can be written approximately as
\begin{equation}
{\sigma}_{2 \to 6} = \int_{2 m_{\pi}}^{{\rm max}\{m_{X_{3}}\}} \int_{2 m_{\pi}}^{{\rm max}\{m_{X_{4}}\}}
{\sigma}_{2 \to 4}(...,m_{X_{3}},m_{X_{4}})\,
f_{M}(m_{X_{3}})\, f_{M}(m_{X_{4}}) \,dm_{X_{3}}\, dm_{X_{4}}
\,.
\label{4pi_amplitude}
\end{equation}
Here we use for the calculation of the decay processes
$M \to \pi^{+} \pi^{-}$ with $M =\sigma, \rho$
the spectral function 
%
\begin{equation}
f_{M}(m_{X_{i}}) = 
\left( 1-\dfrac{4 m_{\pi}^{2}}{m_{X_{i}}^{2}} \right)^{n/2}
\frac{\frac{2}{\pi}{m_{M}^{2}} \Gamma_{M,tot}}{(m_{X_{i}}^{2}-m_{M}^{2})^{2} + m_{M}^{2} \Gamma_{M,tot}^{2}}\,N_{I}\,,
\label{spectral_function}
\end{equation}
where $i  = 3, 4$.
In (\ref{spectral_function}) $n = 3$, $N_{I} = 1$ 
for $\rho$ meson 
and $n = 1$, $N_{I} = \frac{2}{3}$ for $\sigma$ meson, respectively.
The quantity $\left( 1-4 m_{\pi}^{2}/m_{X_{i}}^{2} \right)^{n/2}$
smoothly decreases the spectral function when approaching the $\pi^{+}\pi^{-}$ threshold,
$m_{X_{i}} \to 2 m_{\pi}$.

\subsection{$pp \to pp \sigma \sigma$}
\label{sec:section_f0f0}
Here we discuss the exclusive production 
of $\sigma \sigma \equiv f_{0}(500) f_{0}(500)$ pairs in proton-proton collisions,
%
\begin{eqnarray}
p(p_{a},\lambda_{a}) + p(p_{b},\lambda_{b}) \to
p(p_{1},\lambda_{1}) + \sigma(p_{3}) + \sigma(p_{4}) + p(p_{2},\lambda_{2}) \,,
\label{2to4_reaction_f0f0}
\end{eqnarray}
where $p_{a,b}$, $p_{1,2}$ and $\lambda_{a,b}$, $\lambda_{1,2} = \pm \frac{1}{2}$ 
denote the four-momenta and helicities of the protons
and $p_{3,4}$ denote the four-momenta of the mesons, respectively.

The diagram for the $\sigma \sigma$ production with an intermediate $\sigma$ meson
is shown in Fig.~\ref{fig:4pi_diagram1}~(left diagram).
The amplitude for this process can be written as the following sum:
\begin{eqnarray}
{\cal M}^{(\sigma\mathrm{-exchange})}_{pp \to pp \sigma\sigma} &=&
{\cal M}^{(\Pom \Pom \to \sigma\sigma)} +
{\cal M}^{(\Pom f_{2 \Reg} \to \sigma\sigma)} +
{\cal M}^{(f_{2 \Reg} \Pom \to \sigma\sigma)} +
{\cal M}^{(f_{2 \Reg} f_{2 \Reg} \to \sigma\sigma)}\,.
\label{2to4_reaction_SS}
\end{eqnarray}
For instance, the $\Pom\Pom$-exchange amplitude can be written as
\begin{eqnarray}
{\cal M}^{(\Pom \Pom \to \sigma\sigma)} =
{\cal M}^{({\hat{t}})}_{\lambda_{a} \lambda_{b} \to \lambda_{1} \lambda_{2} \sigma\sigma}+
{\cal M}^{({\hat{u}})}_{\lambda_{a} \lambda_{b} \to \lambda_{1} \lambda_{2} \sigma\sigma}
\label{2to4_reaction_SS_pompom}
\end{eqnarray}
with the $\hat{t}$- and $\hat{u}$-channel amplitudes
\begin{equation}
\begin{split}
& {\cal M}^{({\hat{t}})}_{\lambda_{a} \lambda_{b} \to \lambda_{1} \lambda_{2} \sigma\sigma} 
= \\
& \quad (-i)
\bar{u}(p_{1}, \lambda_{1}) 
i\Gamma^{(\Pom pp)}_{\mu_{1} \nu_{1}}(p_{1},p_{a}) 
u(p_{a}, \lambda_{a})\,
i\Delta^{(\Pom)\, \mu_{1} \nu_{1}, \alpha_{1} \beta_{1}}(s_{13},t_{1}) \,
i\Gamma^{(\Pom \sigma\sigma)}_{\alpha_{1} \beta_{1}}(p_{t},-p_{3}) \,
i\Delta^{(\sigma)}(p_{t}) \\
& \quad \times  i\Gamma^{(\Pom \sigma\sigma)}_{\alpha_{2} \beta_{2}}(p_{4},p_{t})\,
i\Delta^{(\Pom)\, \alpha_{2} \beta_{2}, \mu_{2} \nu_{2}}(s_{24},t_{2}) \,
\bar{u}(p_{2}, \lambda_{2}) 
i\Gamma^{(\Pom pp)}_{\mu_{2} \nu_{2}}(p_{2},p_{b}) 
u(p_{b}, \lambda_{b}) \,,
\end{split}
\label{amplitude_t}
\end{equation}
\begin{equation} 
\begin{split}
& {\cal M}^{({\hat{u}})}_{\lambda_{a} \lambda_{b} \to \lambda_{1} \lambda_{2} \sigma\sigma} 
= \\ 
& \quad (-i)\,
\bar{u}(p_{1}, \lambda_{1}) 
i\Gamma^{(\Pom pp)}_{\mu_{1} \nu_{1}}(p_{1},p_{a}) 
u(p_{a}, \lambda_{a}) \,
i\Delta^{(\Pom)\, \mu_{1} \nu_{1}, \alpha_{1} \beta_{1}}(s_{14},t_{1}) \,
i\Gamma^{(\Pom \sigma\sigma)}_{\alpha_{1} \beta_{1}}(p_{4},p_{u}) 
\,i\Delta^{(\sigma)}(p_{u})  \\
& \quad \times  
i\Gamma^{(\Pom \sigma\sigma)}_{\alpha_{2} \beta_{2}}(p_{u},-p_{3})\, 
i\Delta^{(\Pom)\, \alpha_{2} \beta_{2}, \mu_{2} \nu_{2}}(s_{23},t_{2}) \,
\bar{u}(p_{2}, \lambda_{2}) 
i\Gamma^{(\Pom pp)}_{\mu_{2} \nu_{2}}(p_{2},p_{b}) 
u(p_{b}, \lambda_{b}) \,,
\end{split}
\label{amplitude_u}
\end{equation}
where $p_{t} = p_{a} - p_{1} - p_{3}$,
$p_{u} = p_{4} - p_{a} + p_{1}$, $s_{ij} = (p_{i} + p_{j})^{2}$,
%
$t_1 = (p_{1} - p_{a})^{2}$, $t_2 = (p_{2} - p_{b})^{2}$.
Here $\Delta^{(\Pom)}$ and $\Gamma^{(\Pom pp)}$ 
denote the effective propagator and proton vertex function,
respectively, for the tensorial pomeron.
The effective propagators and vertex functions for the tensorial 
pomeron/reggeon exchanges respect the standard crossing and 
charge-conjugation relations of Quantum Field Theory.
For the explicit expressions of these terms see 
Sect.~3 of \cite{Ewerz:2013kda}.
We assume that $\Gamma^{(\Pom \sigma \sigma)}$ 
has the same form as $\Gamma^{(\Pom \pi \pi)}$
(see (3.45) of \cite{Ewerz:2013kda})
but with the $\Pom \sigma \sigma$ coupling constant
$g_{\Pom \sigma \sigma}$
instead of the $\Pom \pi \pi$ one $2 \beta_{\Pom \pi \pi}$.
The scalar meson propagator $\Delta^{(\sigma)}$
is taken as in (4.7) and (4.8) of \cite{Lebiedowicz:2016ioh}
with the running (energy-dependent) width.
In a similar way the $\Pom f_{2 \Reg}$, $f_{2 \Reg} \Pom$ 
and $f_{2 \Reg} f_{2 \Reg}$ amplitudes can be written.
For the $f_{2 \Reg} \sigma \sigma$ vertex our ansatz 
is as for $f_{2 \Reg} \pi \pi$ in (3.53) of \cite{Ewerz:2013kda}
but with $g_{f_{2 \Reg} \pi \pi}$ replaced by $g_{f_{2 \Reg} \sigma \sigma}$.

In the high-energy small-angle approximation 
we can write the $2 \to 4$ amplitude (\ref{2to4_reaction_SS}) as
\begin{equation}
\begin{split}
& {\cal M}^{(\sigma\mathrm{-exchange})}_{\lambda_{a}\lambda_{b}\to\lambda_{1}\lambda_{2}\sigma\sigma}
\simeq\;
2 (p_1 + p_a)_{\mu_{1}} (p_1 + p_a)_{\nu_{1}}\, 
\delta_{\lambda_{1} \lambda_{a}} \, F_{1}(t_{1}) \,F_{M}(t_{1})\\
& \quad \times 
\bigg\{
{V}^{\mu_{1} \nu_{1}}(s_{13}, t_{1}, p_{t}, -p_{3})\;
\Delta^{(\sigma)}(p_{t})\;
{V}^{\mu_{2} \nu_{2}}(s_{24}, t_{2}, p_{t}, p_{4}) 
\, \left[ \hat{F}_{\sigma}(p_{t}^{2}) \right]^{2}\\
& \qquad +
{V}^{\mu_{1} \nu_{1}}(s_{14}, t_{1}, p_{u}, p_{4})\;
\Delta^{(\sigma)}(p_{u})\;
{V}^{\mu_{2} \nu_{2}}(s_{23}, t_{2}, p_{u}, -p_{3}) 
\, \left[ \hat{F}_{\sigma}(p_{u}^{2}) \right]^{2}
\bigg\}   \\
& \quad \times 2 (p_2 + p_b)_{\mu_{2}} (p_2 + p_b)_{\nu_{2}}\, 
\delta_{\lambda_{2} \lambda_{b}} \, F_{1}(t_{2}) \,F_{M}(t_{2})\,.
\end{split}
\label{amplitude_approx_f0f0_sigma}
\end{equation}
The function ${V}_{\mu \nu}$ has the form ($M_{0} \equiv 1$~GeV)
\begin{equation}
\begin{split}
{V}_{\mu \nu}(s,t,k_{2},k_{1})= &
(k_{1}+k_{2})_{\mu}(k_{1}+k_{2})_{\nu}\;
\frac{1}{4s}\bigg[
3 \beta_{\Pom NN} \, g_{\Pom \sigma \sigma} 
(- i s \alpha'_{\Pom})^{\alpha_{\Pom}(t)-1} 
\\
& + \frac{1}{2 M_{0}^{2}} g_{f_{2 \Reg}pp} \, g_{f_{2 \Reg} \sigma \sigma}
(- i s \alpha'_{f_{2 \Reg}})^{\alpha_{f_{2 \Reg}}(t) -1} 
\bigg]  
\,,
\end{split}
\label{tensorial_function_aux}
\end{equation}
where $\beta_{\Pom NN}$ = 1.87~GeV$^{-1}$ and $g_{f_{2 \Reg} pp}$ = 11.04
from (6.53) and (6.55) of \cite{Ewerz:2013kda}, respectively.

If the $\sigma$ meson has substantial gluon content 
or some $q\bar{q}q\bar{q}$ component its coupling to $\Pom$ and $f_{2 \Reg}$
may be larger than for the pion.
To illustrate effects of this possibility we take in the calculation 
two sets of the coupling constants
%
\begin{eqnarray}
&&{\rm set\;A}: \;g_{\Pom \sigma \sigma} = 2 \beta_{\Pom \pi\pi}\,, \quad g_{f_{2 \Reg} \sigma \sigma} = g_{f_{2 \Reg} \pi\pi} \,,
\label{couplings_setA} \\
&&{\rm set\;B}: \;g_{\Pom \sigma \sigma} = 4 \beta_{\Pom \pi\pi}\,, \quad g_{f_{2 \Reg} \sigma \sigma} = 2 g_{f_{2 \Reg} \pi\pi} \,,
\label{couplings_setB}
\end{eqnarray}
where $\beta_{\Pom \pi\pi}$ = 1.76~GeV$^{-1}$ and $g_{f_{2 \Reg} \pi\pi}$ = 9.30
from (7.15) and (7.16) of \cite{Ewerz:2013kda}, respectively.

The form of the off-shell meson form factor $\hat{F}_{\sigma}(k^{2})$ 
in (\ref{amplitude_approx_f0f0_sigma}) 
and of the analogous form factor for $\rho$ mesons
$\hat{F}_{\rho}(k^{2})$ (see the next section) is unknown.
We write generically $\hat{F}_{M}(k^{2})$ $(M = \sigma, \rho)$
for these form factors which we
normalize to unity at the on-shell point,
$\hat{F}_{M}(m_{M}^{2}) = 1$, and parametrise here in two ways:
%
\begin{eqnarray} 
&&\hat{F}_{M}(k^{2})=
\exp\left(\frac{k^{2}-m_{M}^{2}}{\Lambda^{2}_{off,E}}\right) \,,
\label{off-shell_form_factors_exp} \\
&&\hat{F}_{M}(k^{2})=
\dfrac{\Lambda^{2}_{off,Mp} - m_{M}^{2}}{\Lambda^{2}_{off,Mp} - k^{2}} \,, \quad   \Lambda_{off,Mp}>m_{M} \,.
\label{off-shell_form_factors_mon} 
\end{eqnarray}
The cut-off parameters $\Lambda_{off,E}$ for the exponential form or 
$\Lambda_{off,Mp}$ for the monopole form of the form factors
can be adjusted to experimental data.

A factor $\frac{1}{2}$ due to the identity of the two $\sigma$ mesons in
the final state has to be taken into account in the phase-space
integration in (\ref{4pi_amplitude}).
\subsection{$pp \to pp \rho \rho$}
\label{sec:section_rho0rho0}

Here we focus on exclusive production of 
$\rho \rho \equiv \rho(770) \rho(770)$ in proton-proton collisions,
see Fig.~\ref{fig:4pi_diagram1}~(right diagram),
\begin{eqnarray}
p(p_{a},\lambda_{a}) + p(p_{b},\lambda_{b}) \to
p(p_{1},\lambda_{1}) + \rho(p_{3},\lambda_{3}) + \rho(p_{4},\lambda_{4}) + p(p_{2},\lambda_{2}) \,,
\label{2to4_reaction_rhorho}
\end{eqnarray}
where $p_{3,4}$ and $\lambda_{3,4} = 0, \pm 1$ 
denote the four-momenta and helicities of the $\rho$ mesons, respectively.
We write the amplitude as
\begin{equation}
\begin{split}
{\cal M}_{\lambda_{a}\lambda_{b}\to\lambda_{1}\lambda_{2}\rho\rho} = 
\left(\epsilon^{(\rho)}_{\rho_{3}}(\lambda_{3})\right)^*
\left(\epsilon^{(\rho)}_{\rho_{4}}(\lambda_{4})\right)^*
{\cal M}^{\rho_{3} \rho_{4}}_{\lambda_{a}\lambda_{b}\to\lambda_{1}\lambda_{2}\rho\rho}
\,,
\end{split}
\label{amplitude_rhorho}
\end{equation}
where $\epsilon^{(\rho)}_{\rho}(\lambda)$ are the polarisation vectors of the $\rho$ meson.

Then, with the expressions for the propagators, vertices, and form factors,
from \cite{Ewerz:2013kda} ${\cal M}^{\rho_{3} \rho_{4}}$ can be written
in the high-energy approximation as
%
%
\begin{equation}
\begin{split}
& {\cal M}^{(\rho\mathrm{-exchange})\,\rho_{3} \rho_{4}}_{\lambda_{a}\lambda_{b}\to\lambda_{1}\lambda_{2}\rho\rho}
\simeq \;
2 (p_1 + p_a)_{\mu_{1}} (p_1 + p_a)_{\nu_{1}}\, 
\delta_{\lambda_{1} \lambda_{a}} \, F_{1}(t_{1}) \,F_{M}(t_{1})\\
& \times 
\bigg\{
{V}^{\rho_{3} \rho_{1} \mu_{1} \nu_{1}}(s_{13}, t_{1}, p_{t}, p_{3})\;
\Delta^{(\rho)}_{\rho_{1}\rho_{2}}(p_{t})\;
{V}^{\rho_{4} \rho_{2} \mu_{2} \nu_{2}}(s_{24}, t_{2}, -p_{t}, p_{4}) 
\, \left[ \hat{F}_{\rho}(p_{t}^{2}) \right]^{2}\\
& \quad \quad \; +
{V}^{\rho_{4} \rho_{1} \mu_{1} \nu_{1}}(s_{14}, t_{1}, -p_{u}, p_{4})\;
\Delta^{(\rho)}_{\rho_{1}\rho_{2}}(p_{u})\;
{V}^{\rho_{3} \rho_{2} \mu_{2} \nu_{2}}(s_{23}, t_{2}, p_{u}, p_{3}) 
\, \left[ \hat{F}_{\rho}(p_{u}^{2}) \right]^{2}
\bigg\}   \\
& \times 2 (p_2 + p_b)_{\mu_{2}} (p_2 + p_b)_{\nu_{2}}\, 
\delta_{\lambda_{2} \lambda_{b}} \, F_{1}(t_{2}) \,F_{M}(t_{2}) \,,
\end{split}
\label{amplitude_approx}
\end{equation}
where ${V}_{\mu \nu \kappa \lambda}$ reads as
%
\begin{equation}
\begin{split}
{V}_{\mu \nu \kappa \lambda}(s,t,k_{2},k_{1})= &
2 \Gamma_{\mu \nu \kappa \lambda}^{(0)}(k_{1},k_{2})
\frac{1}{4s}\bigg[
3 \beta_{\Pom NN} \, a_{\Pom \rho\rho} 
(- i s \alpha'_{\Pom})^{\alpha_{\Pom}(t)-1} 
\\
& \qquad \qquad \qquad \quad \; + \frac{1}{M_{0}} g_{f_{2 \Reg}pp} \, a_{f_{2 \Reg} \rho \rho}
(- i s \alpha'_{f_{2 \Reg}})^{\alpha_{f_{2 \Reg}}(t) -1} 
\bigg]  \\
& -\Gamma_{\mu \nu \kappa \lambda}^{(2)}(k_{1},k_{2})
\frac{1}{4s}\bigg[
3 \beta_{\Pom NN} \, b_{\Pom \rho \rho} 
(- i s \alpha'_{\Pom} )^{\alpha_{\Pom}(t)-1} 
\\
& \qquad \qquad  \qquad \quad+ \frac{1}{M_{0}} g_{f_{2 \Reg}pp} \, b_{f_{2 \Reg} \rho \rho}
(- i s \alpha'_{f_{2 \Reg}})^{\alpha_{f_{2 \Reg}}(t) -1}  
\bigg] \,.
\end{split}
\label{tensorial_function_aux2}
\end{equation}
The explicit tensorial functions 
$\Gamma_{\mu \nu \kappa \lambda}^{(i)}(k_{1},k_{2})$, 
$i$ = 0, 2, are given in Ref.~\cite{Ewerz:2013kda}, 
see formulae (3.18) and (3.19), respectively.
In our calculations
the parameter set~A of coupling constants 
$a$ and $b$ from \cite{Lebiedowicz:2014bea} was used, see Eq.~(2.15) there.

We consider in (\ref{2to4_reaction_rhorho}) unpolarised protons
in the initial state and no observation of polarizations in the final state.
In the following we are mostly interested in the invariant mass distributions
of the $4\pi$ system and in distributions of the parent $\rho$ mesons.
Therefore, we have to insert in (\ref{4pi_amplitude})
the cross section $\sigma_{2 \to 4}$ summed over the $\rho$ meson polarizations.
The spin sum for a $\rho$ meson of momentum $k$ and squared mass $k^{2}=m_{X}^{2}$ is
\begin{equation}
\begin{split}
\sum_{\lambda = 0, \pm 1}
\epsilon^{(\rho)\,\mu}(\lambda)
\left(\epsilon^{(\rho)\,\nu}(\lambda)\right)^* =
-g^{\mu \nu} + \dfrac{k^{\mu}k^{\nu}}{m_{X}^{2}}
\,.
\end{split}
\label{spinsum}
\end{equation}
But the $k^{\mu}k^{\nu}$ terms do not contribute since we have the relations
\begin{equation}
\begin{split}
p_{3\, \rho_{3}} {\cal M}^{\rho_{3} \rho_{4}}_{\lambda_{a}\lambda_{b}\to\lambda_{1}\lambda_{2}\rho\rho} =0\,, \qquad
p_{4\, \rho_{4}} {\cal M}^{\rho_{3} \rho_{4}}_{\lambda_{a}\lambda_{b}\to\lambda_{1}\lambda_{2}\rho\rho} =0
\,.
\end{split}
\label{spinsum_aux}
\end{equation}
These follow from the properties of $\Gamma^{(0,2)}_{\mu\nu\kappa\lambda}$
in (\ref{tensorial_function_aux2}); see (3.21) of \cite{Ewerz:2013kda}.

Taking also into account the statistical factor $\frac{1}{2}$ due to the identity
of the two $\rho$ mesons we get for the amplitudes squared 
(to be inserted in $\sigma_{2 \to 4}$ in (\ref{4pi_amplitude}))
\begin{equation}
\begin{split}
&\frac{1}{2} \frac{1}{4} \sum_{\rm{spins}}
\Big|\left(\epsilon^{(\rho)}_{\rho_{3}}(\lambda_{3})\right)^*
\left(\epsilon^{(\rho)}_{\rho_{4}}(\lambda_{4})\right)^*
{\cal M}^{\rho_{3} \rho_{4}}_{\lambda_{a}\lambda_{b}\to\lambda_{1}\lambda_{2}\rho\rho}
\Big|^{2}
\\
&=\frac{1}{8} \sum_{\lambda_{a},\lambda_{b},\lambda_{1},\lambda_{2}}
\left({\cal M}^{\sigma_{3} \sigma_{4}}_{\lambda_{a}\lambda_{b}\to\lambda_{1}\lambda_{2}\rho\rho}\right)^{*}
{\cal M}^{\rho_{3} \rho_{4}}_{\lambda_{a}\lambda_{b}\to\lambda_{1}\lambda_{2}\rho\rho}\,
g_{\sigma_{3}\rho_{3}} \,g_{\sigma_{4}\rho_{4}}
\,.
\end{split}
\label{amplitude_squared_rhorho}
\end{equation}

So far we have treated the exchanged mesonic object for
$\rho^0 \rho^0$ production as spin-1 particle.
However, we should take into account the fact that the exchanged 
intermediate object is not a simple meson but may correspond 
to a whole family of daughter exchanges, that is, 
the reggeization of the intermediate $\rho$ meson is necessary.
For related works, where this effect was included in practical calculations, 
see e.g. \cite{Cisek:2011vt,Lebiedowicz:2013ika}.
The ``reggeization'' of the amplitude given in Eq.~(\ref{amplitude_approx})
is included here for $\sqrt{s_{34}} \geqslant 2 m_{\rho}$, 
only approximately, by replacing the $\rho$ propagator 
both in the $\hat{t}$- and $\hat{u}$-channel amplitudes by
%
\begin{eqnarray}
&&\Delta^{(\rho)}_{\rho_{1}\rho_{2}}(p) \to
\Delta^{(\rho)}_{\rho_{1}\rho_{2}}(p) 
\left( \frac{s_{34}}{s_{0}} \right)^{\alpha_{\rho}(p^{2})-1} \,,
\label{reggeization}
\end{eqnarray}
where we take $s_{0} = 4 m_{\rho}^{2}$ and 
$\alpha_{\rho}(p^{2}) = 0.5 + 0.9 \,t$
with the momentum transfer $t = p^{2}$.


To give the full physical amplitudes 
we should add absorptive corrections to the Born amplitudes 
(\ref{amplitude_approx_f0f0_sigma}) and (\ref{amplitude_approx})
for the $pp \to pp \sigma \sigma$ and $pp \to pp \rho \rho$ reactions, respectively.
For the details how to include the $pp$-rescattering corrections 
in the eikonal approximation for the four-body reaction
see Sect.~3.3 of \cite{Lebiedowicz:2014bea}.

\section{Preliminary results}
\label{sec:section_4}

In this section we wish to present first results for the
$pp \to pp \sigma \sigma$ and $pp \to pp \rho \rho$  processes
corresponding to the diagrams shown in Fig.~\ref{fig:4pi_diagram1}.

\begin{figure}[!ht]
\includegraphics[width=0.45\textwidth]{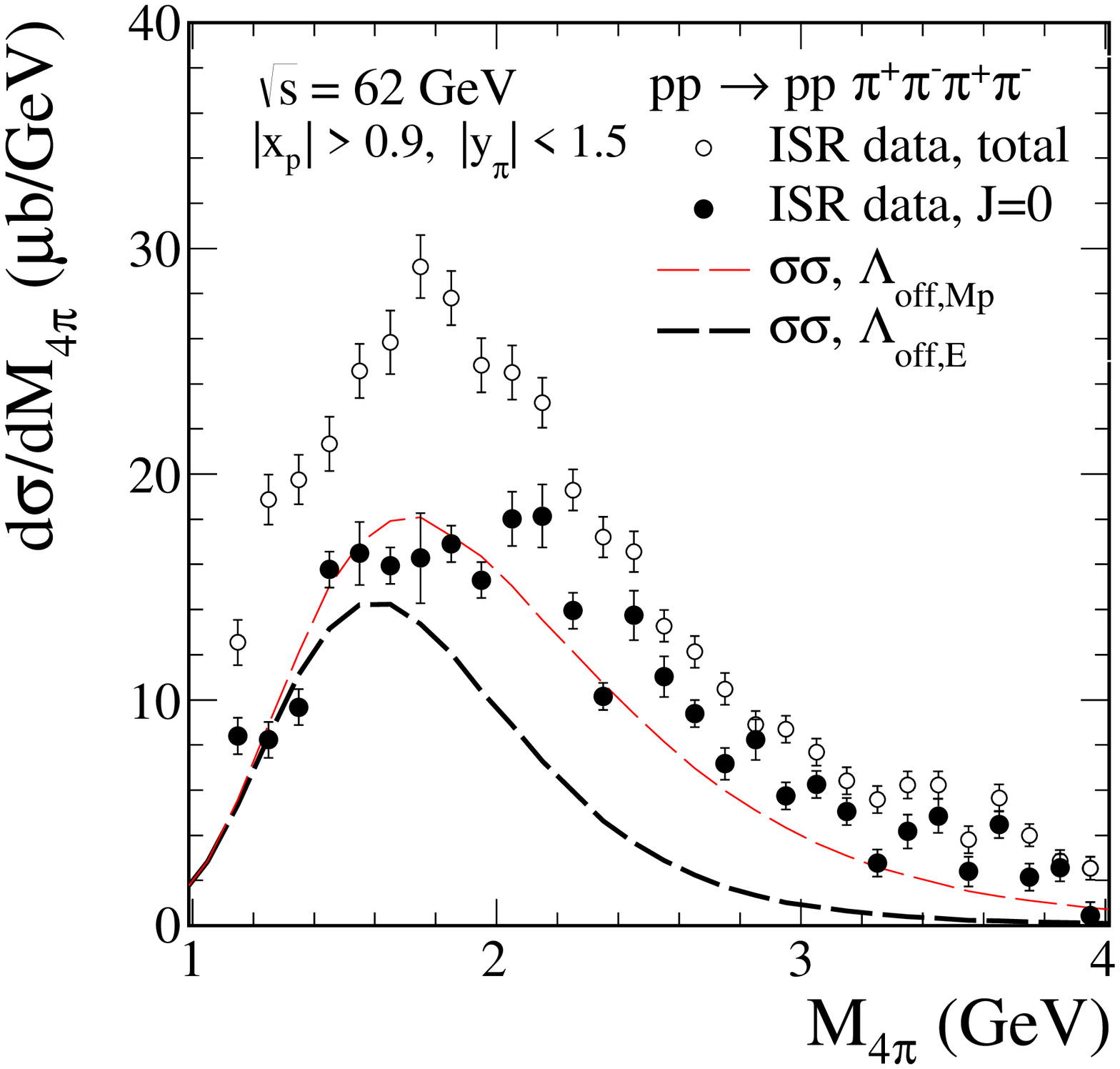}
\includegraphics[width=0.45\textwidth]{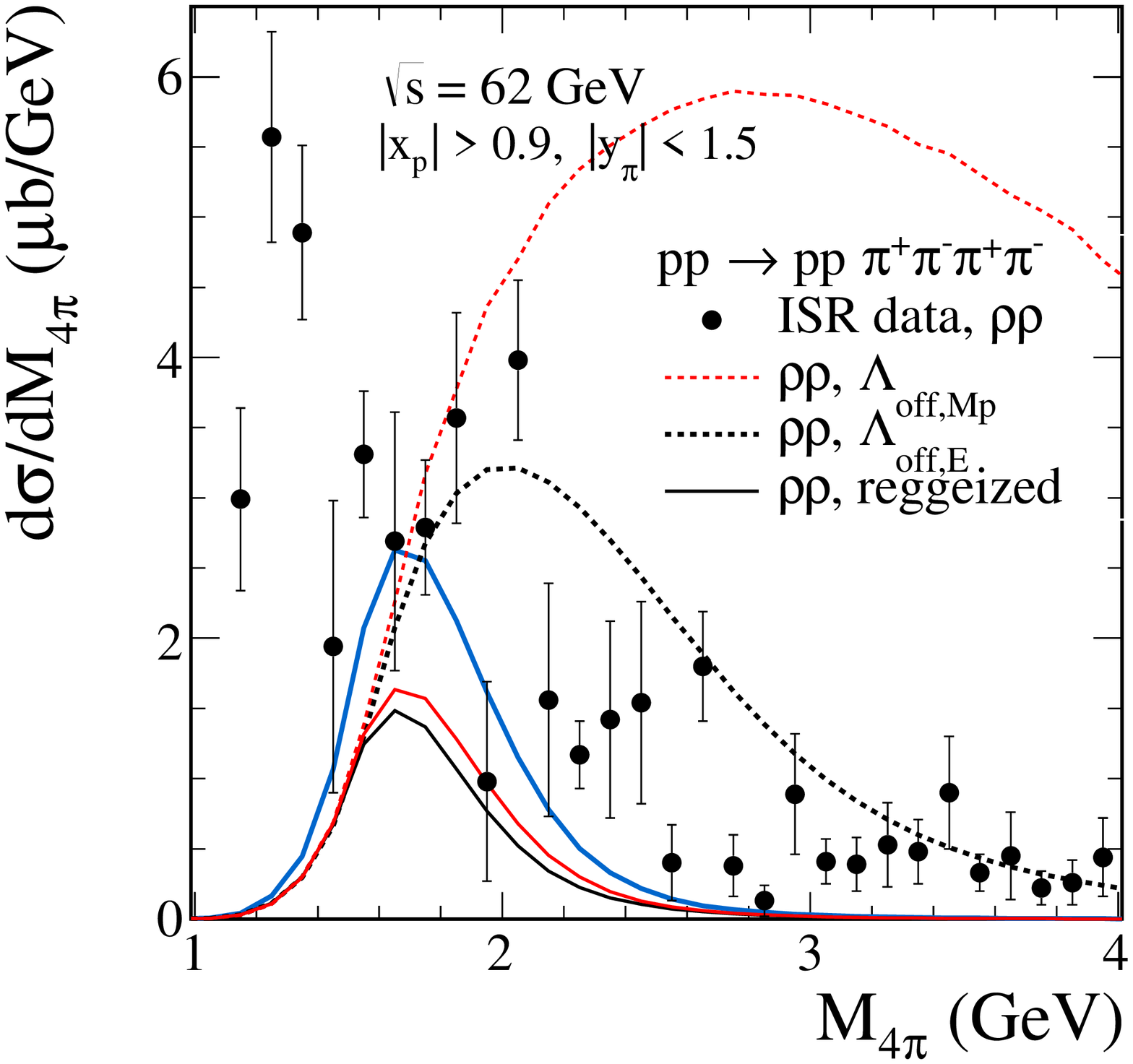}
  \caption{\label{fig:M4pi}
  \small
Invariant mass distributions for the central $\pi^{+}\pi^{-}\pi^{+}\pi^{-}$ system
compared to the CERN-ISR data \cite{Breakstone:1993ku} at $\sqrt{s} = 62$~GeV.
In the left panel the lines represent results for the 
$\sigma\sigma$ contribution only and 
with the enhanced pomeron/reggeon-$\sigma$-$\sigma$ couplings (\ref{couplings_setB}).
We used two forms of the off-shell meson form factor,
the exponential form (\ref{off-shell_form_factors_exp}) 
with $\Lambda_{off,E} = 1.6$~GeV (the black lines)
and the monopole form (\ref{off-shell_form_factors_mon}) 
with $\Lambda_{off,Mp} = 1.6$~GeV (the red thin lines).
In the right panel the lines represent results for the $\rho\rho$ contribution
without (the dotted lines) and with (the solid lines) the inclusion 
of the intermediate $\rho$ meson reggeization.
For comparison, the upper blue solid line was obtained 
with the monopole form factor and $\Lambda_{off,Mp} = 1.8$~GeV.
The absorption effects were included here.
}
\end{figure}

We start from a discussion of the $\pi^{+}\pi^{-}\pi^{+}\pi^{-}$
invariant mass distribution.
In Fig.~\ref{fig:M4pi}
we compare the $\sigma \sigma$- and $\rho \rho$-contributions
to the CERN-ISR data \cite{Breakstone:1993ku} at $\sqrt{s} = 62$~GeV.
Here, the four pions are restricted to lie in the rapidity region $|y_{\pi}| < 1.5$
and the cut \footnote{The Feynman-$x$ variable was defined as 
$x_{p} = 2 p_{z,p}/\sqrt{s}$
in the center-of-mass frame with $p_{z,p}$ the longitudinal momentum 
of the outgoing proton.} 
$|x_{p}| > 0.9$ is imposed on the outgoing protons.
In Ref.~\cite{Breakstone:1993ku} five contributions to the four-pion spectrum
were identified. A $4 \pi$ phase-space term
with total angular momentum $J = 0$,
two $\rho \pi \pi$ terms with $J = 0$ and $J = 2$,
and two $\rho \rho$ terms with $J = 0$ and $J = 2$.
In the following we will compare the $4 \pi$ phase-space term
with our $\sigma \sigma$ result,
the $\rho \rho$ terms with our $\rho \rho$ result.
The theoretical results correspond to the calculation
including absorptive corrections related to the $pp$ nonperturbative interaction
in the initial and final state.
The ratio of full and Born cross sections $\langle S^{2}\rangle$
(the gap survival factor) is approximately $\langle S^{2}\rangle = 0.4$.
In our calculation both the $\Pom \Pom$ 
and the $\Pom f_{2 \Reg}$, $f_{2 \Reg} \Pom$, $f_{2 \Reg} f_{2 \Reg}$ exchanges were included.
At the ISR energy the $f_{2 \Reg}$ exchanges, including their interference terms
with the $\Pom \Pom$ one, give about $50\%$ to the total cross section.

In the left panel of Fig.~\ref{fig:M4pi} we compare our $\sigma \sigma$ contribution
assuming the coupling constants (\ref{couplings_setB}) 
with the $4 \pi$ ($J=0$, phase space) ISR data (marked as full data points).
We present results for two different forms of off-shell meson form factor,
the exponential type (\ref{off-shell_form_factors_exp}), $\Lambda_{off,E} = 1.6$~GeV,
and the monopole type (\ref{off-shell_form_factors_mon}), $\Lambda_{off,Mp} = 1.6$~GeV, 
see the black lower line and the red upper line, respectively.
There is quite a good agreement between our $\sigma \sigma$ result
with a monopole form factor and the $4 \pi$ ($J=0$, phase space) data.
Note that this implies that the set~B of $\Pom \sigma \sigma$
and $f_{2 \Reg} \sigma \sigma$ couplings, which are larger than
the corresponding pion couplings, seems to be preferred.

In the right panel of Fig.~\ref{fig:M4pi} we compare our result 
for the $\rho\rho$ contribution to corresponding ISR data
\footnote{Here we plotted the sum of the experimental cross section
$\sigma = \sigma_{J=0} + \sigma_{J=2}$
for the $J=0$ and $J=2$ $\rho\rho$ terms
(see Figs.~3c and 3e in \cite{Breakstone:1993ku}, respectively)
and the corresponding error is approximated as 
$\delta \sigma = \sqrt{\delta_{J=0}^{2} + \delta_{J=2}^{2}}$.}.
In the calculation of the $\rho \rho$ contribution we take into account the 
intermediate $\rho$ meson reggeization.
The reggeization leads to an extra strong damping of the large $M_{4 \pi}$ cross section.
The effect of reggeization is expected only 
when the separation in rapidity between the two produced resonances is large.
We will return to this issue in the further part of this section.
We note that our model is able to give
a qualitative account of the ISR $\rho \rho$ data for $M_{4 \pi} \gtrsim 1.4$~GeV
within the large experimental errors.

The total $4 \pi$ experimental data (marked as open data points in the left panel
of Fig.~\ref{fig:M4pi}) are also shown for comparison.
In Ref.~\cite{Breakstone:1993ku} an integrated (total) cross section of 46~$\mu b$
at $\sqrt{s} = 62$~GeV was estimated.
There are other processes besides the ones of (\ref{2to6_reaction})
contributing to the $4 \pi$ final state, such as resonance production
shown in the diagrams in Fig.~\ref{fig:4pi_diagram2} of Appendix~\ref{sec:diagram2}.
Thus, the $\sigma \sigma$ and the $\rho \rho$ contributions considered here
should not be expected to fit the ISR data precisely.
In addition, the ISR $4 \pi$ data includes also 
a large $\rho^{0} \pi^{+} \pi^{-}$ ($J=0$ and $J=2$) component 
(see Figs.~3b and 3d of \cite{Breakstone:1993ku})
with an enhancement in the $J=2$ term which was interpreted there as a $f_{2}(1720)$ state.
Also the $\rho^{0} \rho^{0}$ ($J=2$) term
indicates a signal of $f_{2}(1270)$ state; see Fig.~3e of \cite{Breakstone:1993ku}.
Therefore, a consistent model for the resonance and continuum contributions,
including the interference effects between them, 
would be required to better describe the ISR data. 
We leave this interesting problem for future studies.

\begin{figure}[!ht]
\includegraphics[width=0.45\textwidth]{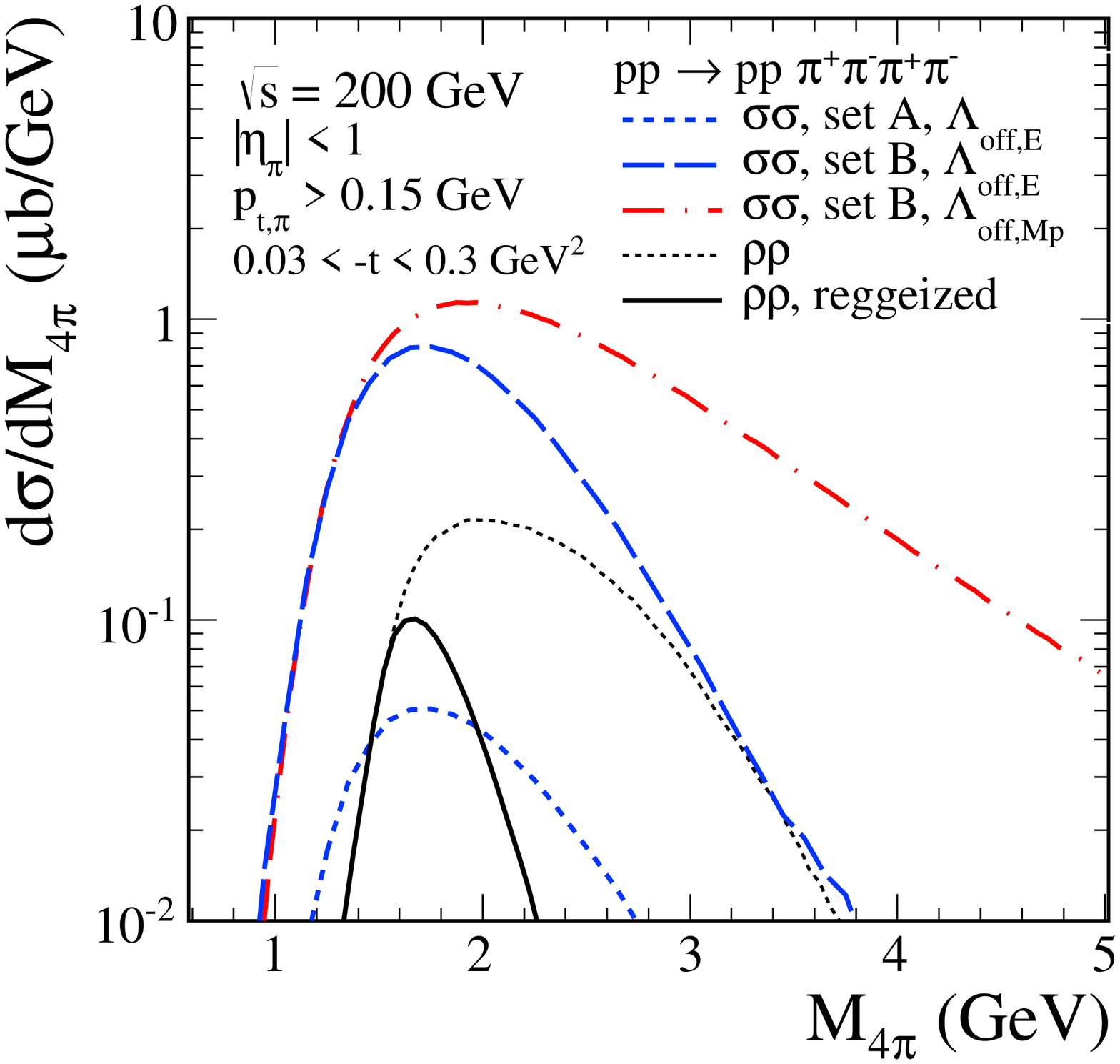}
\includegraphics[width=0.45\textwidth]{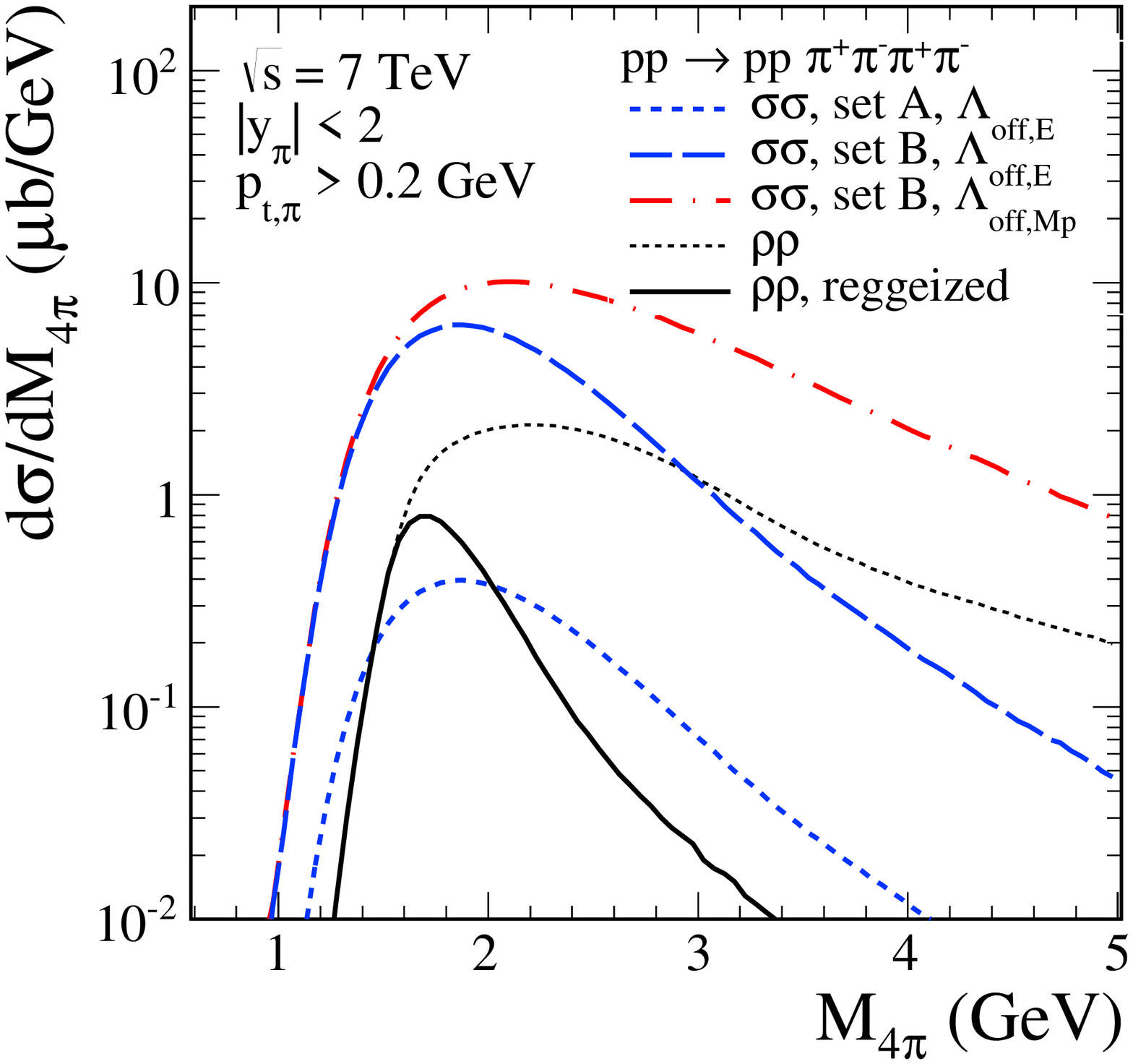}
\includegraphics[width=0.45\textwidth]{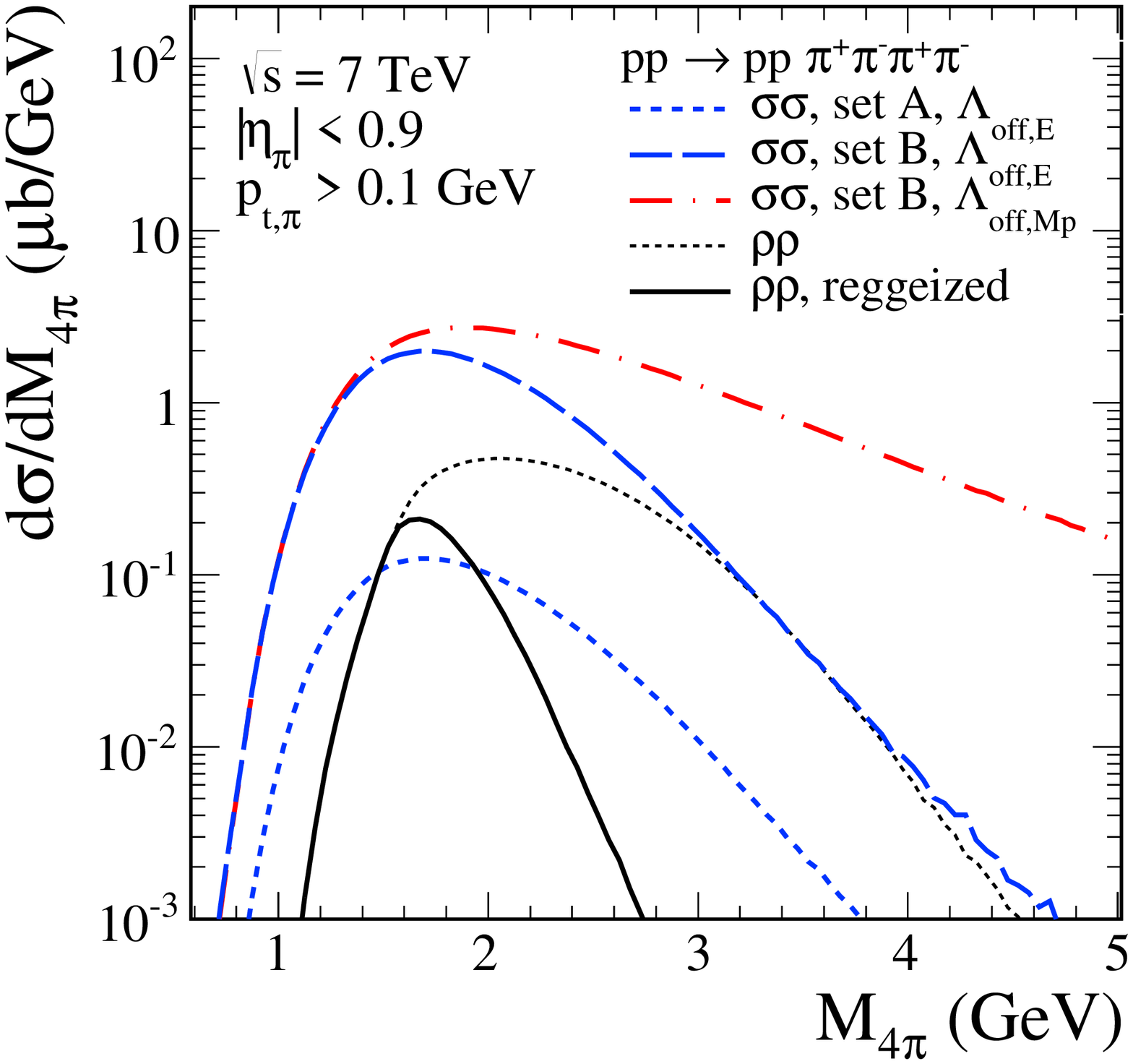}
\includegraphics[width=0.45\textwidth]{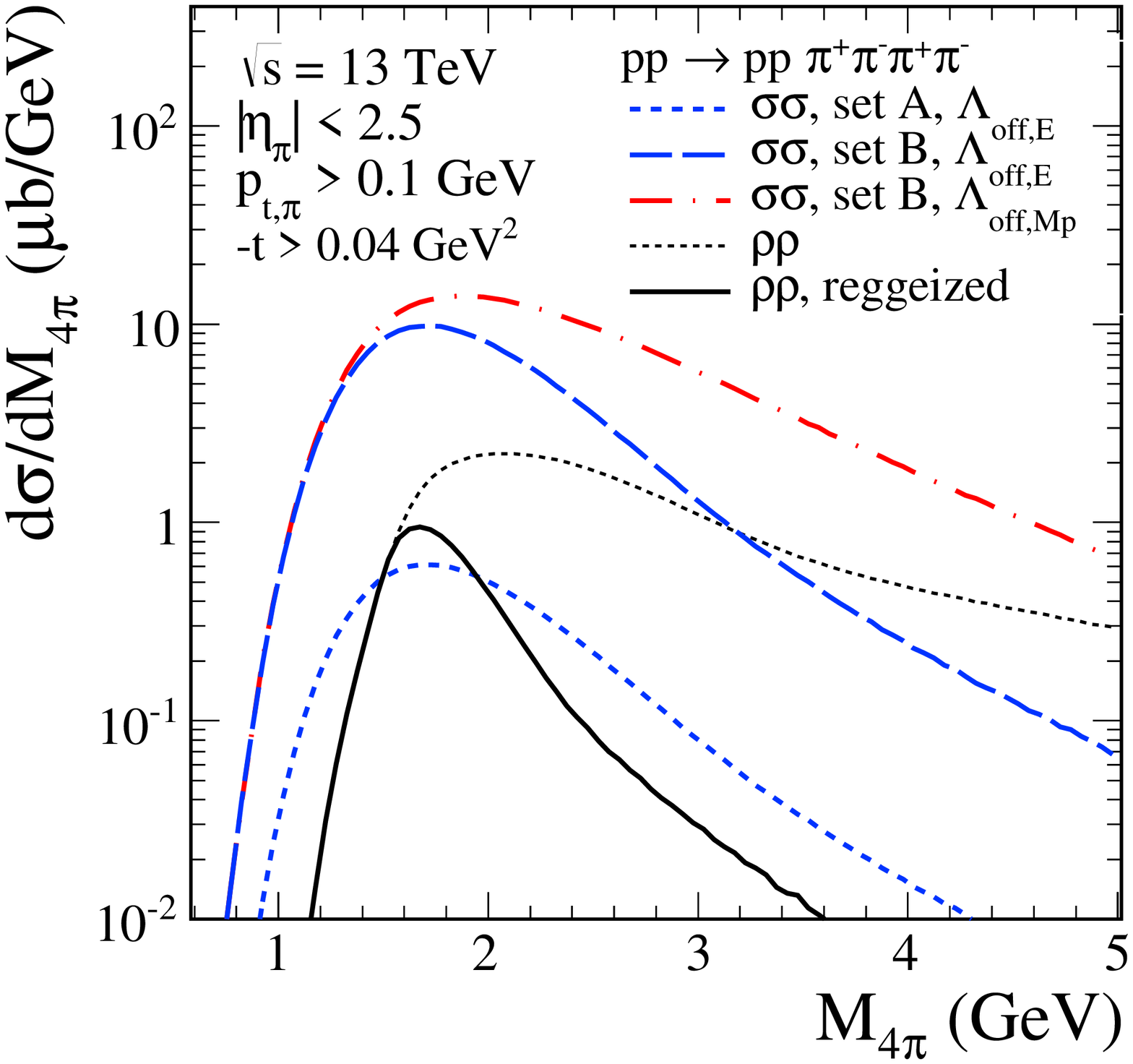}
  \caption{\label{fig:M4pi_PP}
  \small
Four-pion invariant mass distributions for different 
center-of-mass energies $\sqrt{s}$ and experimental kinematical cuts.
The black lines represent results for the $\rho\rho$ contribution,
the blue lines for the $\sigma \sigma$ contribution.
The exponential off-shell meson form factors (\ref{off-shell_form_factors_exp})
with $\Lambda_{off,E} = 1.6$~GeV were used.
For the case of $\sigma \sigma$ contribution only
the red dot-dashed line was obtained 
with the monopole form factor and $\Lambda_{off,Mp} = 1.6$~GeV.
The absorption effects were included here.
}
\end{figure}

In Fig.~\ref{fig:M4pi_PP} we show our preliminary four-pion invariant mass
distributions for experimental cuts relevant for the RHIC and LHC experiments.
In the calculation of the $\sigma \sigma$ and the $\rho \rho$ contributions
the pomeron and $f_{2 \Reg}$ exchanges were included.
Imposing limitations on pion (pseudo)rapidities, e.g. $|\eta_{\pi}| < 1$,
and going to higher energies strongly reduces the role of subleading $f_{2 \Reg}$ exchanges.
The gap survival factors $\langle S^{2}\rangle$
estimated within the eikonal approximation are 
0.30, 0.21, 0.23 for $\sqrt{s}$ = 0.2, 7, 13~TeV, respectively.
In the case of $\sigma \sigma$ contribution we use 
two sets of the coupling constants;
standard (\ref{couplings_setA}) and enhanced ones (\ref{couplings_setB}).

\begin{figure}[!ht]
\includegraphics[width = 0.45\textwidth]{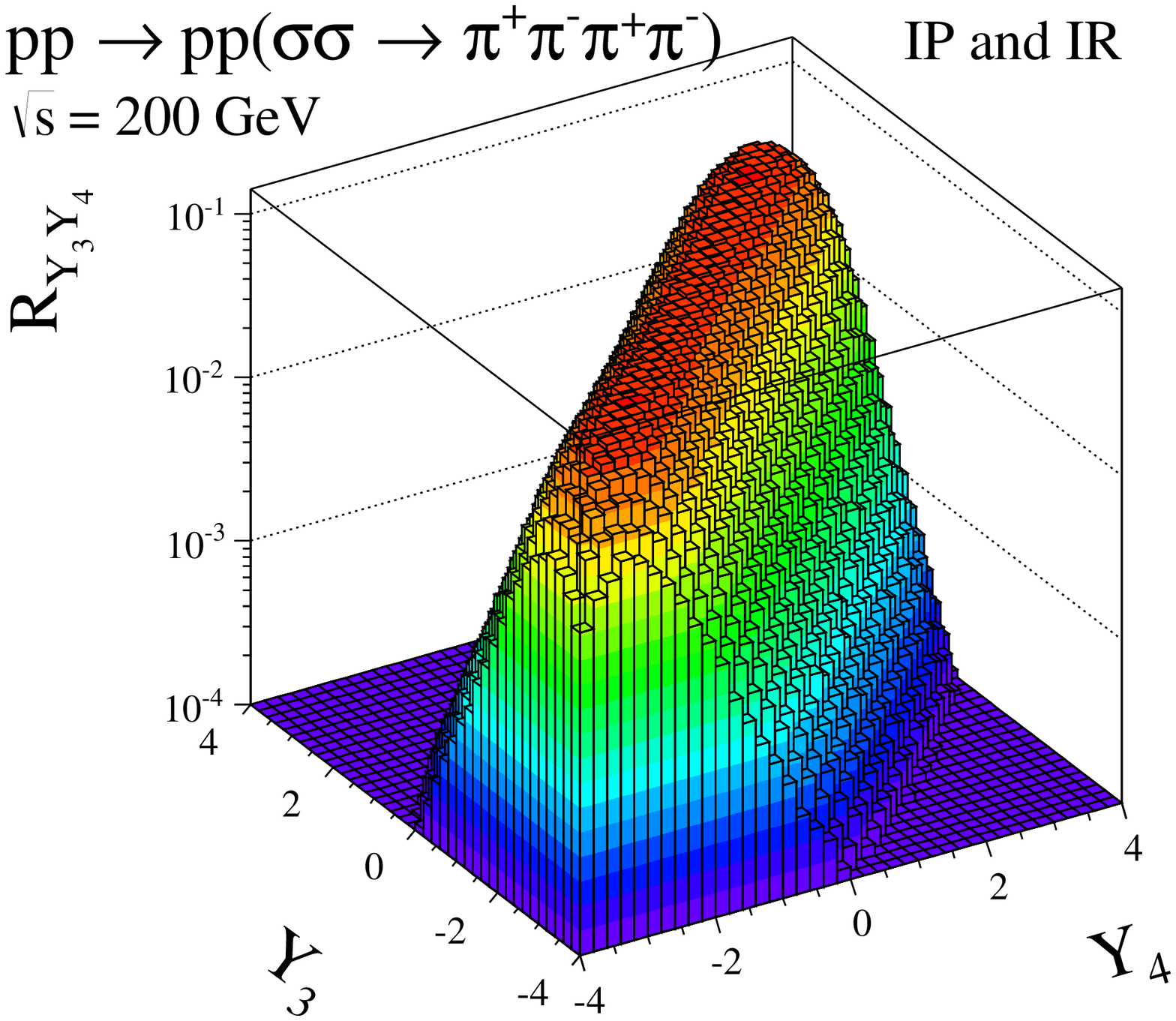}\\
\includegraphics[width = 0.45\textwidth]{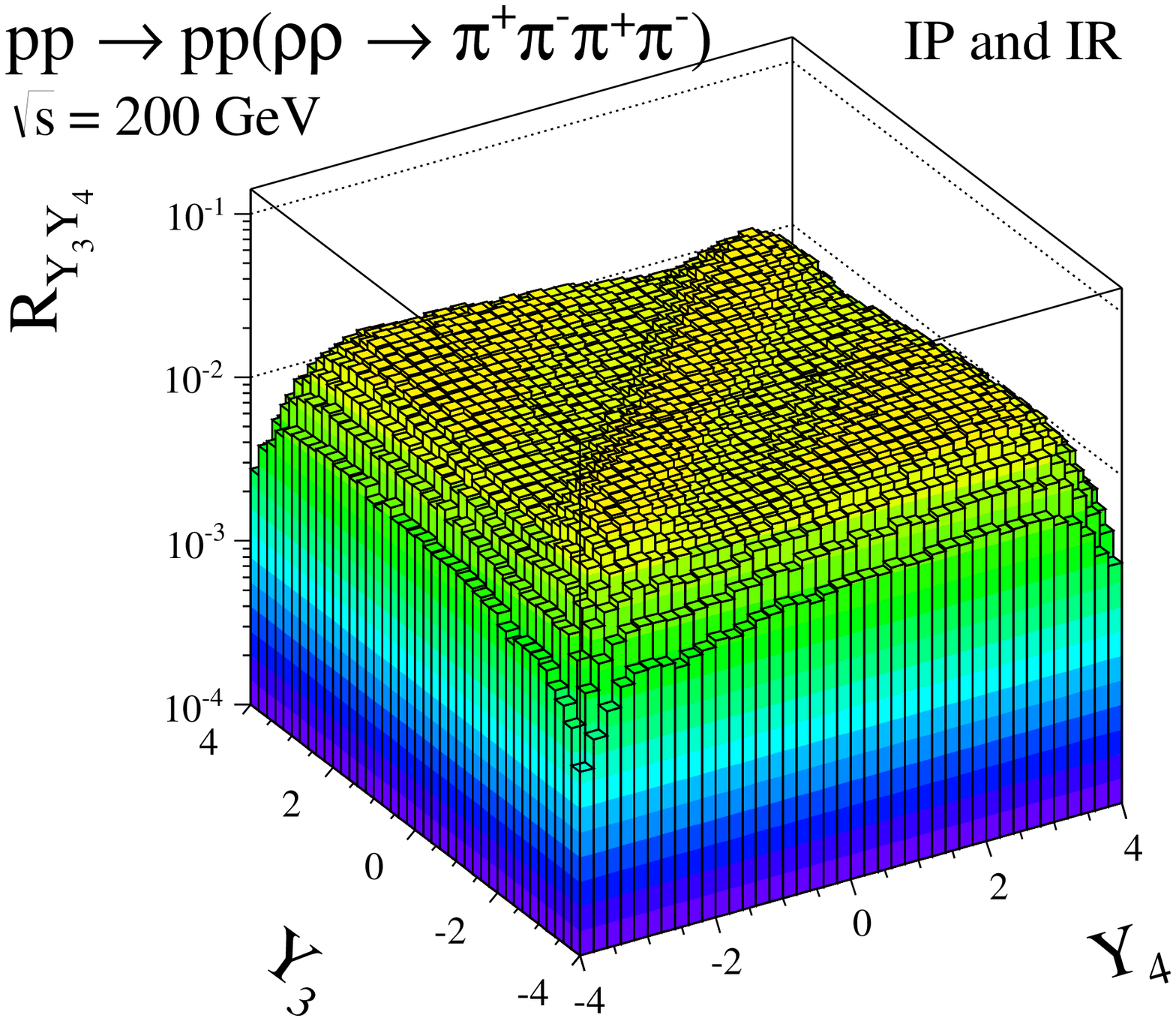}
\includegraphics[width = 0.45\textwidth]{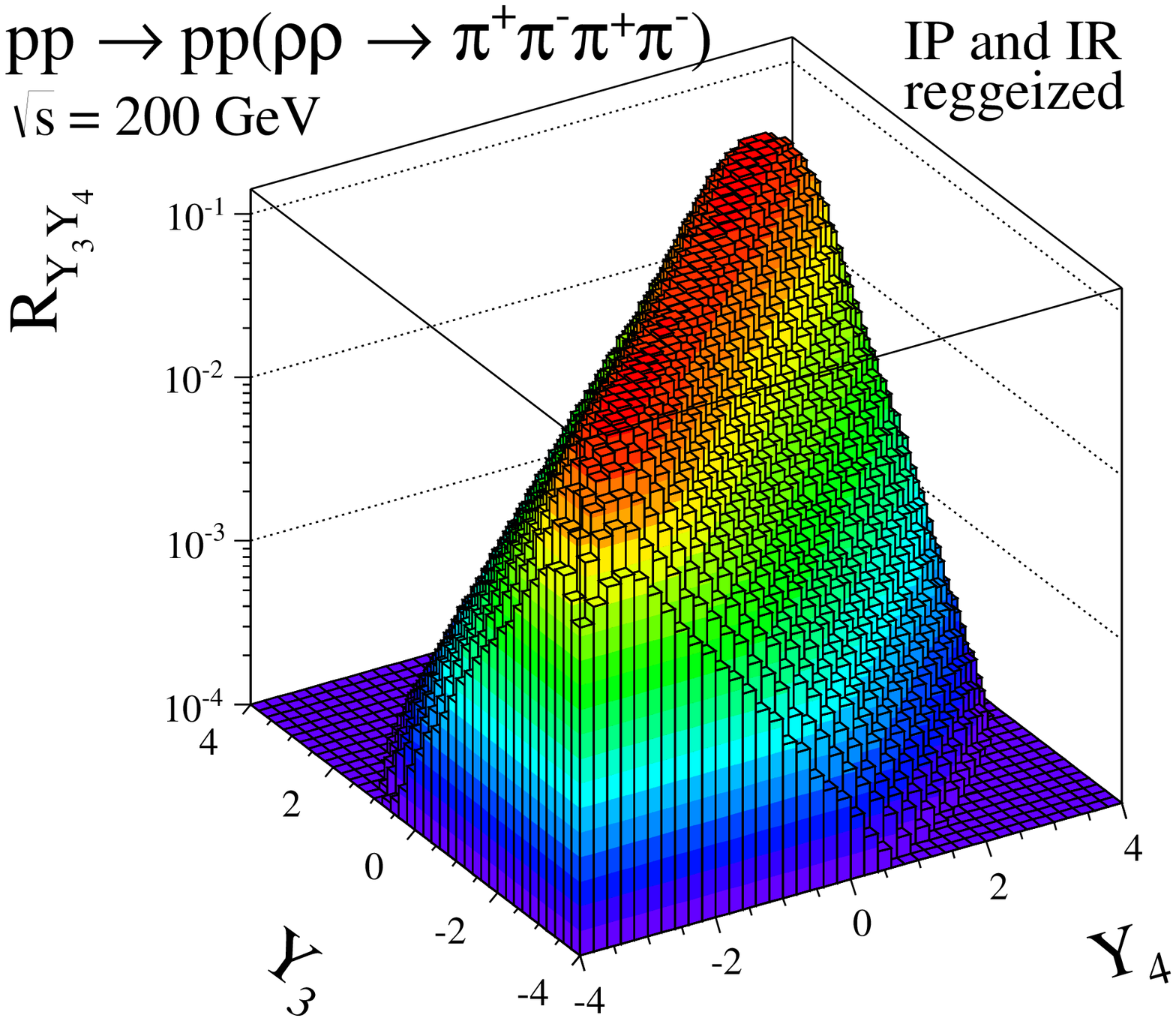}
  \caption{\label{fig:ratios}
  \small
The distributions in $({\rm Y}_{3},{\rm Y}_{4})$ space
for the reaction $pp \to pp \pi^{+}\pi^{-}\pi^{+}\pi^{-}$ 
via fusion of two tensor pomerons and $f_{2 \Reg}$ reggeons
at $\sqrt{s} = 200$~GeV.
Plotted is the ratio 
$R_{{\rm Y}_{3}{\rm Y}_{4}} = \frac{d^{2}\sigma}{d{\rm Y}_{3}d{\rm Y}_{4}} 
/ \int{d{\rm Y}_{3}d{\rm Y}_{4} \frac{d^{2}\sigma}{d{\rm Y}_{3}d{\rm Y}_{4}}}$.
We show the $\sigma \sigma$ contribution (top panel) 
and the $\rho \rho$ contribution (bottom panels)
without (left panel) and with (right panel) the $\rho$ meson
reggeization included.
Here $\Lambda_{off,E} = 1.6$~GeV was used.
}
\end{figure}

The correlation in rapidity of pion pairs $({\rm Y}_{3},{\rm Y}_{4})$
(e.g., ${\rm Y}_3$ means ${\rm Y}_{\pi^{+} \pi^{-}}$
where the pions are produced from a meson decay)
for both the $\sigma\sigma$ and the $\rho\rho$ contributions
is displayed in Fig.~\ref{fig:ratios} for $\sqrt{s} = 200$~GeV.
For the $\sigma\sigma$ contribution, see the top panel,
we observe a strong correlation ${\rm Y}_3 \approx {\rm Y}_4$. 
For the $\rho \rho$ case the $({\rm Y}_3, {\rm Y}_4)$ distribution
extends over a much broader range of ${\rm Y}_3 \ne {\rm Y}_4$
which is due to the exchange of the spin-1 particle.
However, as discussed in the section devoted to the formalism we may
include, at least approximately, the effect of reggeization of 
the intermediate $\rho$ meson.
In the left and right bottom panels we show the results without 
and with the $\rho$ meson reggeization, respectively.
As shown in the right panel this effect becomes crucial 
when the separation in rapidity between the two $\rho$ mesons increases.
After the reggeization is performed the two-dimensional distribution
looks very similar as for the $\sigma \sigma$ case.
The reggeization effect discussed here is also closely related to 
the damping of the four-pion invariant mass distribution discussed
already in Fig.~\ref{fig:M4pi}.

In Table~\ref{tab:table1} we have collected integrated cross sections 
in $\mu b$ with different experimental cuts
for the exclusive $\pi^{+}\pi^{-} \pi^{+}\pi^{-}$ production
including only the contributions shown in Fig.~\ref{fig:4pi_diagram1}.
The collected results were obtained in the calculations with 
the tensor pomeron and reggeon exchanges.
In the calculations the off-shell-meson form factor 
(\ref{off-shell_form_factors_exp}) with $\Lambda_{off,E} = 1.6$~GeV was used. 
No absorption effects were included in the quoted numbers.
The full cross section can be obtained by multiplying the Born cross section 
by the corresponding gap survival factor $\langle S^{2}\rangle$.
These factors depend on the kinematic cuts and are
0.40 (ISR), 0.46 (STAR, lower $|t|$), 0.30 (STAR, higher $|t|$), 
0.21 ($\sqrt{s}$ = 7~TeV), 0.19 ($\sqrt{s}$ = 13~TeV),
0.23 ($\sqrt{s}$ = 13~TeV, with cuts on $|t|$).
\begin{table}[]
\centering
\caption{The integrated ``Born level'' (no gap survival factors) cross sections in $\mu b$ 
for the central exclusive $\pi^{+}\pi^{-}\pi^{+}\pi^{-}$ production
in $pp$ collisions via the $\sigma\sigma$ and $\rho\rho$ mechanisms
given in Fig.~\ref{fig:4pi_diagram1}
for some typical experimental cuts.
The $\sigma\sigma$ contribution was calculated using the coupling
constants (\ref{couplings_setB})
while the $\rho \rho$ contribution without
and with (in the parentheses) the inclusion of the intermediate $\rho$ meson reggeization.
}
\label{tab:table1}
\begin{tabular}{ll|c|c|c}
\cline{3-4}
                       &  & \multicolumn{2}{l|}{Cross sections in $\mu b$} &  \\ \cline{1-4}
\multicolumn{1}{|l|}{$\sqrt{s}$, TeV} & Cuts 
&     $\sigma\sigma$      &    $\rho\rho$       &  \\ \cline{1-4}
\multicolumn{1}{|l|}{0.062} &  $|y_{\pi}| < 1.5$, $|x_{p}| > 0.9$ 
& \;\;37.92  &   \;\;10.66 (2.17)         &  \\ 
\multicolumn{1}{|l|}{0.2} &  $|\eta_{\pi}| < 1$, $p_{t, \pi} > 0.15$~GeV, 
0.005~$< -t <$~0.03~GeV$^{2}$
&  \;\;\;0.30         &    \;\;\;0.10 (0.02)       &  \\ 
\multicolumn{1}{|l|}{0.2} &  $|\eta_{\pi}| < 1$, $p_{t, \pi} > 0.15$~GeV, 
0.03~$< -t <$~0.3~GeV$^{2}$
&  \;\;\;2.94         &    \;\;\;0.88 (0.17)       &  \\ 
\multicolumn{1}{|l|}{7} & $|\eta_{\pi}| < 0.9$, $p_{t, \pi} > 0.1$~GeV
&  \;\;10.40         & \;\;\;2.79 (0.53)         &  \\ 
\multicolumn{1}{|l|}{7} &  $|y_{\pi}| < 2$, $p_{t, \pi} > 0.2$~GeV
&  \;\;34.88         &   \;\;17.94 (2.20)        &  \\ 
\multicolumn{1}{|l|}{13} &  $|\eta_{\pi}| < 1$, $p_{t, \pi} > 0.1$~GeV
&  \;\;16.18       &   \;\;\;3.56 (0.72)        &  \\ 
\multicolumn{1}{|l|}{13} &  $|\eta_{\pi}| < 2.5$, $p_{t, \pi} > 0.1$~GeV
&  120.06      &   \;\;45.58 (6.21)        &  \\ 
\multicolumn{1}{|l|}{13} &  $|\eta_{\pi}| < 2.5$, $p_{t, \pi} > 0.1$~GeV,
$-t >$~0.04~GeV$^{2}$
&  \;\;47.52      &   \;\;18.08 (2.44)        &  \\ 
\cline{1-4}
\end{tabular}
\end{table}
The cross sections for the $\rho \rho$ final state found here
are more than three orders of magnitude larger than
the cross sections for the $\gamma \gamma \to \rho \rho$
and double scattering mechanisms considered recently in \cite{Goncalves:2016ybl}.

\begin{figure}[!ht]
\includegraphics[width=0.45\textwidth]{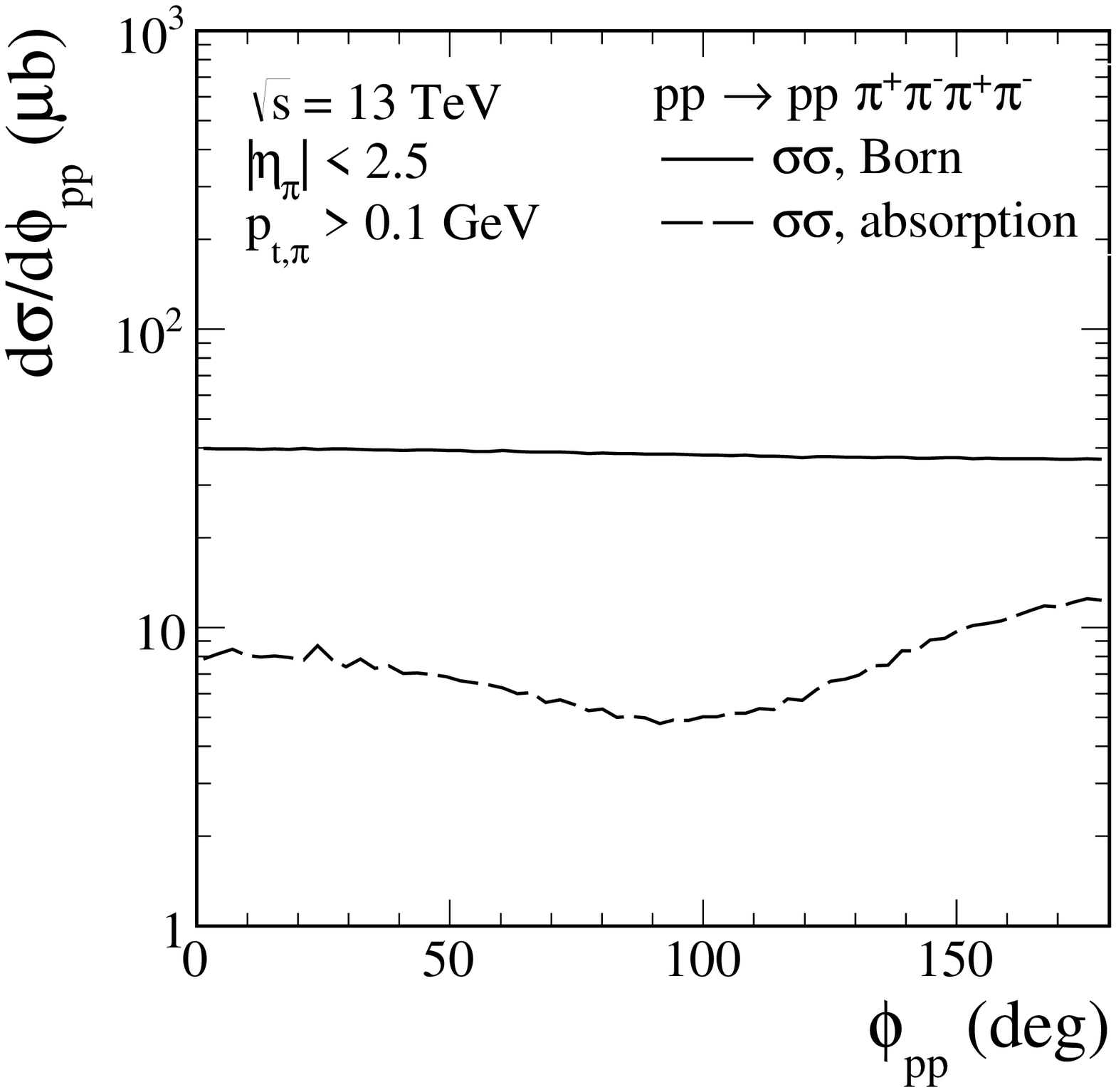}
\includegraphics[width=0.45\textwidth]{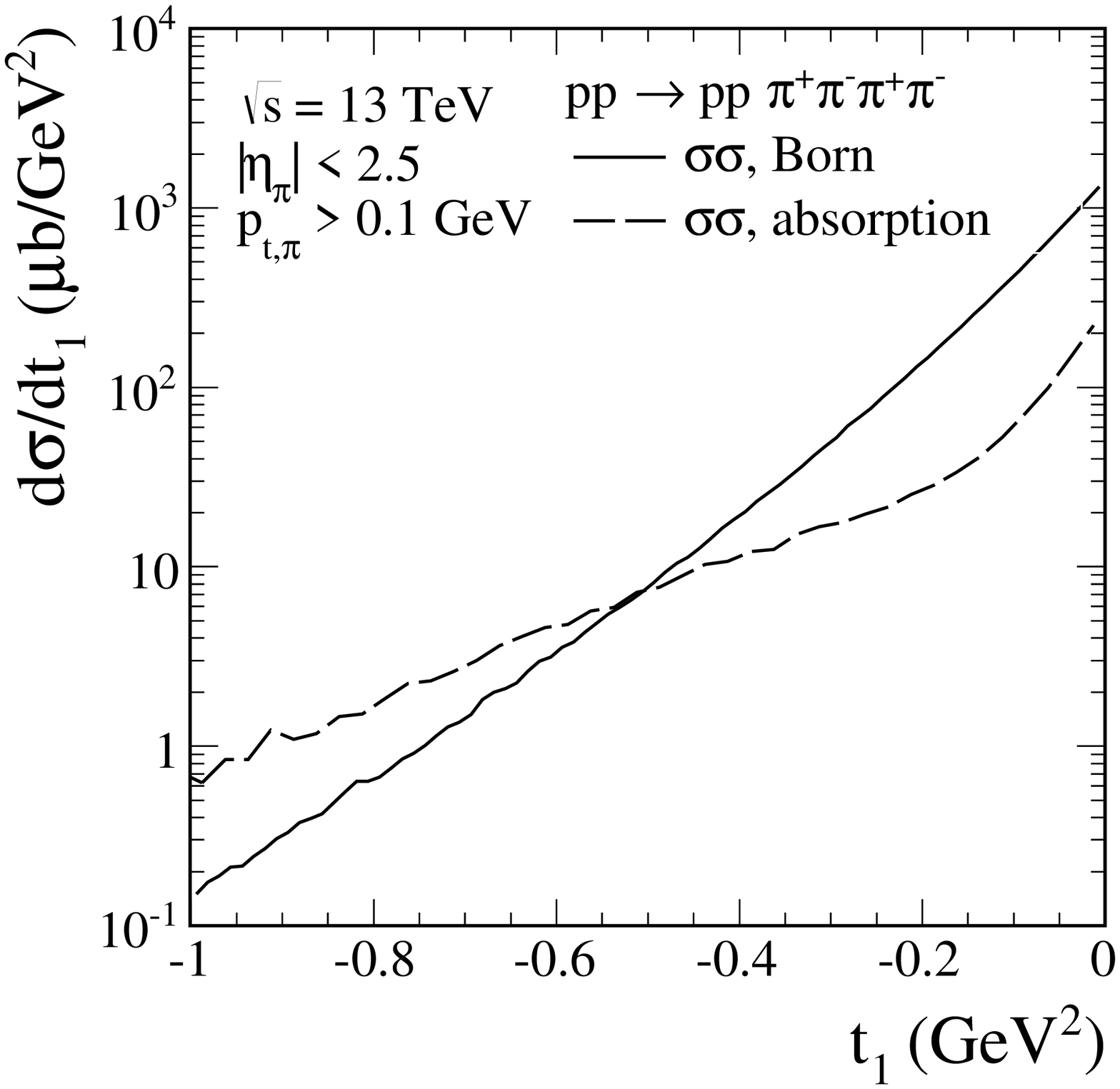}
  \caption{\label{fig:abs}
  \small
Distributions in proton-proton relative azimuthal angle (the left panel) and
in four-momentum squared of one of the protons (the right panel) 
for the central $\sigma\sigma \to \pi^{+}\pi^{-}\pi^{+}\pi^{-}$
contribution at $\sqrt{s} = 13$~TeV with the kinematical cuts 
specified in the legend.
The solid line corresponds to the Born calculations
and the long-dashed line corresponds to the result including
the $pp$ absorptive corrections.
Here the enhanced pomeron/reggeon-$\sigma$-$\sigma$ couplings (\ref{couplings_setB}) 
and the exponential form of off-shell meson form factor (\ref{off-shell_form_factors_exp}) 
with $\Lambda_{off,E} = 1.6$~GeV were used.
}
\end{figure}

Finally, in Fig.~\ref{fig:abs}, we discuss some observables
which are very sensitive to the absorptive corrections.
Quite a different pattern can be seen for the Born case and 
for the case with absorption included. 
The absorptive corrections lead to significant modification of the shape
of the $\phi_{pp}$ distribution
($\phi_{pp}$ is the azimuthal angle between the $p_{t}$ vectors of the outgoing protons)
and lead to an increase of the cross section
for the proton four-momentum transfer $t$ = $t_{1}$ = $t_{2}$ at large $|t|$.
This effect could be verified in future experiments when both protons 
are measured which should be possible for ATLAS-ALFA \cite{Staszewski:2011bg} and CMS-TOTEM.

\section{Conclusions}

In the present paper we have presented first estimates of 
the contributions with the intermediate 
$f_{0}(500)f_{0}(500)$, $\rho(770) \rho(770)$ resonance pairs
to the reaction $p p \to p p \pi^+ \pi^- \pi^+ \pi^-$
which is being analyzed experimentally by the STAR, ALICE, CMS, 
and ATLAS Collaborations.
The results were obtained within a model 
where the pomeron and $f_{2 \Reg}$ reggeon are treated
as effective tensor exchanges.
The results for processes with the exchange of heavy (compared to pion)
mesons strongly depend on the details of the hadronic form factors.
By comparing the theoretical results and the cross sections
found in the CERN-ISR experiment \cite{Breakstone:1993ku}
we fixed the parameters of the off-shell meson form factor
and the $\Pom \sigma \sigma$ and $f_{2 \Reg} \sigma \sigma$ couplings.
The corresponding values of parameters can be verified
by future experimental results obtained at RHIC and LHC.

We have made estimates of the integrated cross sections 
for different experimental situations as well as shown several
differential distributions.
The pion-pair rapidities of the two $\sigma$ mesons 
are strongly correlated
(${\rm Y}_{3} \approx {\rm Y}_{4}$).
This is due to a strong interference effect between the 
$\hat{t}$- and $\hat{u}$-channel amplitudes.
For the $\rho \rho$ contribution the situation is the following.
If we take for the propagator of the exchanged $\rho$
in Fig.~\ref{fig:4pi_diagram1}, right panel, the standard particle propagator
the ${\rm Y}_{3} - {\rm Y}_{4}$ correlation is very weak.
But for a reggeized $\rho$ propagator we get again a strong 
${\rm Y}_{3} - {\rm Y}_{4}$ correlation,
similar to that found in the $\sigma \sigma$ case; see Fig.~\ref{fig:ratios}.
This is understandable since the reggeization suppresses contributions
where the two produced $\rho$ mesons have large subsystem energies, 
i.e. where there is a large rapidity distance between the two $\rho$ mesons.


We have found in this paper that the diffractive mechanism
in proton-proton collisions considered by us
leads to the cross section for the $\rho \rho$ final state more than 
three orders of magnitude larger than the corresponding cross section
for $\gamma \gamma \to \rho \rho$ and double scattering photon-pomeron
(pomeron-photon) mechanisms considered recently in \cite{Goncalves:2016ybl}.

Closely related to the reaction $pp \to pp \pi^{+} \pi^{-} \pi^{+} \pi^{-}$
studied by us here is the $4 \pi$ production
in ultra-peripheral nucleus-nucleus collisions.
A phenomenological study of the reaction mechanism of $AA \to AA \rho^{0} \rho^{0}$
was performed in Ref.~\cite{KlusekGawenda:2013dka}.
The application of our methods, based on the tensor-pomeron concept,
to collisions involving nuclei is an interesting problem
which goes, however, beyond the scope of the present work.

To summarize: we have given a consistent treatment of 
the $\pi^{+}\pi^{-}\pi^{+}\pi^{-}$ production
via two scalar $\sigma$ mesons and two vector $\rho$ mesons
in an effective field-theoretic approach.
A measurable cross section of order of a few $\mu b$ was obtained 
for the $pp \to pp\pi^{+}\pi^{-}\pi^{+}\pi^{-}$ process
which should give experimentalists
interesting challenges to check and explore it.

\appendix
\section{Four-pion production through $f_{0} \to \sigma \sigma$ and $f_{0} \to \rho \rho$ mechanisms}
\label{sec:diagram2}
Here we discuss the diffractive production of the scalar 
$f_{0}(1370)$, $f_{0}(1500)$, and $f_{0}(1710)$ resonances
decaying at least potentially into the $\pi^+ \pi^- \pi^+ \pi^-$ final state.
We present relevant formulas for the resonance contributions 
that could be used in future analyses. At present a precise calculation
of the resonance contributions to the four-pion channel is not possible 
as some details of the relevant decays are not well understood.

The production and decay properties of the scalar mesons in 
the $4 \pi$ channel, such as the $f_{0}(1370)$ and $f_{0}(1500)$ states,
have been investigated extensively in central diffractive production
by the WA102 Collaboration at $\sqrt{s} = 29.1$~GeV 
\cite{Barberis:1997ve,Barberis:1999wn,Barberis:2000em}
and in $p \bar{p}$ and $\bar{p}n$ annihilations
by the Crystal Barrel Collaboration \cite{Abele:2001js}.
In central production, see Fig.~3 of \cite{Barberis:1999wn}, 
there is a very clear signal from $f_{0}(1500)$ in the $4 \pi$ spectra,
especially in the $\sigma \sigma$ channel,
and some evidence of the broad $f_{0}(1370)$ resonance in 
the $\rho \rho$ channel.
In Fig.~1 of Ref.~\cite{Barberis:2000em}
the $J^{PC}= 0^{++}$ $\rho \rho$ wave from 
the $\pi^+ \pi^- \pi^+ \pi^-$ channel 
in four different $\phi_{pp}$ intervals (each of $45^{o}$) was shown.
A peak below $M_{4 \pi} \simeq 1500$~MeV was clearly seen 
which can be interpreted as the interference effect of the $f_{0}(1370)$ state,
the $f_{0}(1500)$ state 
and the broad $4 \pi$ background.
In principle, also the contributions from the $f_0(1710)$ and $f_0(2020)$ are not excluded.
In Table~1 of \cite{Barberis:2000em} the percentage of each resonance
in three intervals of the so-called ``glueball filter variable'' ($dP_{T}$) was shown.
The idea being that for small differences
in the transverse momentum vectors between the two exchanged ``particles'' 
an enhancement in the production of glueballs relative to $q \bar{q}$ states may occur.
The $dP_{T}$ dependence and the $\phi_{pp}$ distributions presented there are similar to what
was found in the analysis of the $\pi^+ \pi^-$ channel \cite{Barberis:1999cq}.
In Refs.~\cite{Kirk:1999df,Kirk:2000ws} it was shown that 
also the $f_{0}(1710)$ state has a similar behaviour in the azimuthal angle $\phi_{pp}$
and in the $dP_{T}$ variable as the $f_{0}(1500)$ state.
That is, all the undisputed $q \bar{q}$ states are observed to be
suppressed at small $dP_{T}$, but the glueball candidates
$f_{0}(1500)$, $f_{0}(1710)$, together with the enigmatic $f_{0}(980)$, survive.
It was shown in \cite{Barberis:2000em}
that the $f_{0}(1370)$ and $f_{0}(2000)$ have similar $dP_{T}$ and $\phi_{pp}$ dependences.
The fact that $f_{0}(1370)$ and $f_{0}(1500)$ states have different
$dP_{T}$ and $\phi_{pp}$ dependences confirms that these are not simply $J$ dependent phenomena
\footnote{Some essential discrepancy for $f_{0}(1370)$ and $f_{0}(1500)$ states
in the different decay channels was discussed, e.g., 
in Refs.~\cite{Klempt:2007cp,Ochs:2013gi}.
In Ref.~\cite{Janowski:2014ppa}, by using a three-flavor chiral effective approach,
the authors found that $f_{0}(1710)$ is predominantly the gluonic state
and the $\rho \rho \to 4 \pi$ decay channel is strongly suppressed.}.
This is also true for the $J =2$ states, where the $f_{2}(1950)$ state
has different dependences compared to 
the $f_{2}(1270)$ and $f'_{2}(1520)$ states \cite{Barberis:1999cq}.
We wish to emphasize that in \cite{Lebiedowicz:2013ika} 
we obtained a good description of the WA102 experimental distributions
\cite{Barberis:1998ax,Barberis:1999cq,Barberis:1999zh}
for the scalar and pseudoscalar mesons within the framework of the tensor pomeron approach.
The $dP_{T}$ and $\phi_{pp}$ effects can be understood as being due to the fact 
that in general more than one pomeron-pomeron-meson 
coupling structure is possible \cite{Lebiedowicz:2013ika}.
The behaviour of the tensor $f_{2}(1270)$ state was discussed 
recently in \cite{Lebiedowicz:2016ioh}; see Figs.~4 and 5 there.

\begin{figure}
\includegraphics[width=6.5cm]{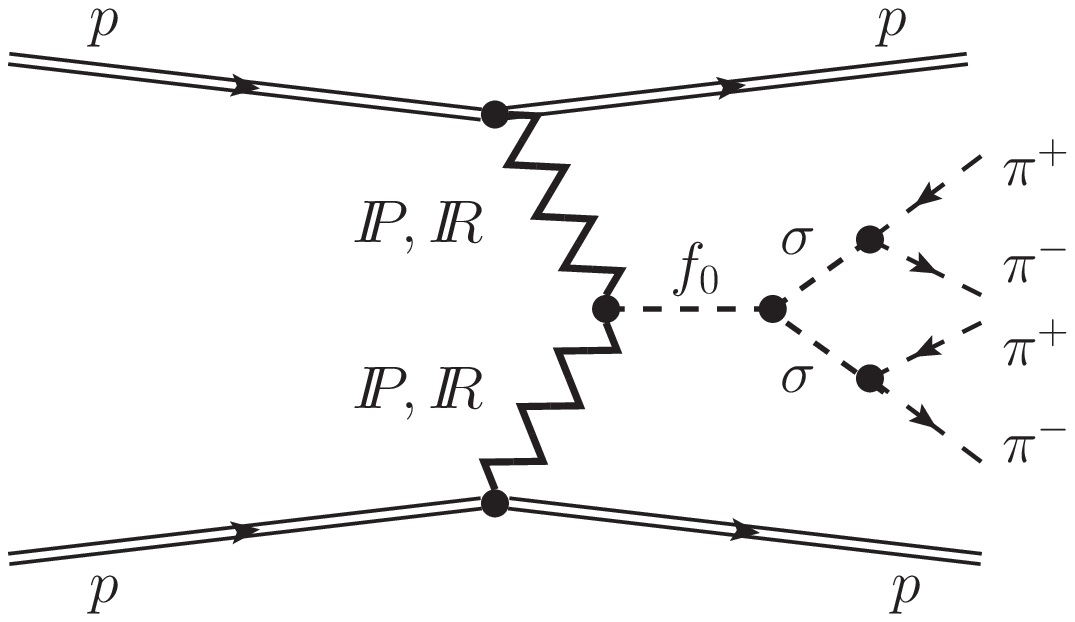}  
\includegraphics[width=6.5cm]{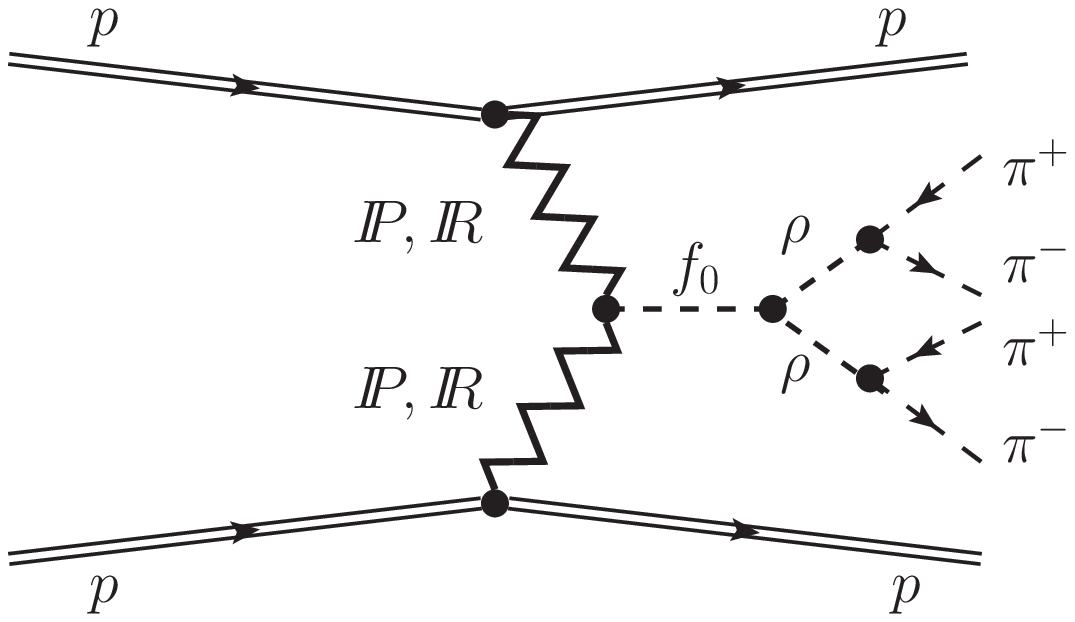}        
  \caption{\label{fig:4pi_diagram2}
  \small The ``Born level'' diagrams for double-pomeron/reggeon
central exclusive production of $\pi^+ \pi^- \pi^+ \pi^-$ 
through $f_{0} \to \sigma \sigma$ (left diagram) and 
$f_{0} \to \rho \rho$ (right diagram)
in proton-proton collisions.
}
\end{figure}

In the following we present our analytic expressions 
for the diagrams of Fig.~\ref{fig:4pi_diagram2}
for $\Pom \Pom$ fusion only. The extension to include
also $\Pom f_{2 \Reg}$, $f_{2 \Reg} \Pom$ and $f_{2 \Reg} f_{2 \Reg}$ fusion
is straightforward.

\subsubsection{$pp \to pp (f_{0} \to \sigma \sigma)$}
\label{sec:section_f0_sigmasigma}

Here we consider the amplitude for the reaction (\ref{2to4_reaction_f0f0})
through an $s$-channel scalar resonance $f_{0} \to \sigma \sigma$;
see Fig.~\ref{fig:4pi_diagram2} (left diagram).
Here $f_{0}$ stands, for 
one of the $f_{0}(1370)$, $f_{0}(1500)$, $f_{0}(1710)$ states.

In the high-energy small-angle approximation we can write this amplitude as
%
\begin{equation}
\begin{split}
& {\cal M}^{(f_{0} \to \sigma\sigma)}_{\lambda_{a}\lambda_{b}\to\lambda_{1}\lambda_{2}\sigma\sigma}
\simeq 3 \beta_{\Pom NN}  \, 2(p_1 + p_a)_{\mu_{1}} (p_1 + p_a)_{\nu_{1}}\, 
\delta_{\lambda_{1} \lambda_{a}}\, F_1(t_1)  \;
\frac{1}{4 s_{1}} (- i s_{1} \alpha'_{\Pom})^{\alpha_{\Pom}(t_{1})-1} \\ 
& \quad \quad
\times 
\Gamma^{(\Pom \Pom f_{0})\,\mu_{1} \nu_{1}, \mu_{2} \nu_{2}}(q_{1},q_{2})\,
\Delta^{(f_{0})}(p_{34})\,
\Gamma^{(f_{0} \sigma \sigma)}(p_{3},p_{4})\\
& \quad \quad\times 
\frac{1}{4 s_{2}} (- i s_{2} \alpha'_{\Pom})^{\alpha_{\Pom}(t_{2})-1}\,
3 \beta_{\Pom NN}  \, 2 (p_2 + p_b)_{\mu_{2}} (p_2 + p_b)_{\nu_{2}}\, 
\delta_{\lambda_{2} \lambda_{b}}\, F_1(t_2) \,,
\end{split}
\label{amplitude_approx_sigmasigma}
\end{equation}
where
$s_{1} = (p_{1} + p_{3} + p_{4})^{2}$,
$s_{2} = (p_{2} + p_{3} + p_{4})^{2}$, 
$q_{1} = p_{a} - p_{1}$, 
$q_{2} = p_{b} - p_{2}$, 
$t_{1} = q_{1}^{2}$, $t_{2} = q_{2}^{2}$,
and
$p_{34} = p_{3} + p_{4}$.

The effective Lagrangians and the vertices for $\Pom \Pom$ fusion into 
an $f_{0}$ meson are discussed in Appendix~A of \cite{Lebiedowicz:2013ika}.
As was shown there the tensorial $\Pom \Pom f_{0}$ vertex 
corresponds to the sum of two lowest values of $(l,S)$, 
that is $(l,S) = (0,0)$ and $(2,2)$
with coupling parameters 
$g_{\Pom \Pom M}'$ and $g_{\Pom \Pom M}''$, respectively.
The vertex, including a form factor, reads then as follows 
($p_{34} = q_{1} + q_{2}$)
\begin{eqnarray}
i\Gamma_{\mu \nu,\kappa \lambda}^{(\Pom \Pom f_{0})} (q_{1},q_{2}) =
\left( i\Gamma_{\mu \nu,\kappa \lambda}'^{(\Pom \Pom f_{0})}\mid_{bare} +
       i\Gamma_{\mu \nu,\kappa \lambda}''^{(\Pom \Pom f_{0})} (q_{1}, q_{2})\mid_{bare} \right)
\tilde{F}^{(\Pom \Pom f_{0})}(q_{1}^{2},q_{2}^{2},p_{34}^{2}) \,;
\label{vertex_pompomS}
\end{eqnarray}
see (A.21) of \cite{Lebiedowicz:2013ika}.
Unfortunately, the pomeron-pomeron-meson form factor
is not well known as it is due to nonperturbative effects related 
to the internal structure of the respective meson.
In practical calculations we take the factorized form
for the $\Pom \Pom f_{0}$ form factor
%
\begin{eqnarray}
\tilde{F}^{(\Pom \Pom f_{0})}(q_{1}^{2},q_{2}^{2},p_{34}^{2}) = 
F_{M}(q_{1}^{2}) F_{M}(q_{2}^{2}) F^{(\Pom \Pom f_{0})}(p_{34}^{2})\,
\label{Fpompommeson}
\end{eqnarray}
normalised to
$\tilde{F}^{(\Pom \Pom f_{0})}(0,0,m_{f_{0}}^{2}) = 1$.
%
We will further set
\begin{eqnarray}
F^{(\Pom \Pom f_{0})}(p_{34}^{2}) = 
\exp{ \left( \frac{-(p_{34}^{2}-m_{f_{0}}^{2})^{2}}{\Lambda_{f_{0}}^{4}} \right)}\,,
\quad \Lambda_{f_{0}} = 1\;{\rm GeV}\,.
\label{Fpompommeson_ff}
\end{eqnarray}

For the $f_{0} \sigma \sigma$ vertex we have 
\begin{eqnarray}
i\Gamma^{(f_{0} \sigma \sigma)}(p_{3},p_{4}) = 
i g_{f_{0} \sigma \sigma} M_{0}\, F^{(f_{0} \sigma \sigma)}(p_{34}^{2})\,,
\end{eqnarray}
where $g_{f_{0} \sigma \sigma}$ is an unknown parameter.
We assume $g_{f_{0} \sigma \sigma}>0$ and 
$F^{(f_{0} \sigma \sigma)}(p_{34}^{2})$ = $F^{(\Pom \Pom f_{0})}(p_{34}^{2})$;
see Eq.~(\ref{Fpompommeson_ff}).
\subsubsection{$pp \to pp (f_{0} \to \rho \rho)$}
\label{sec:section_f0_rho0rho0}

Now we consider the amplitude for the reaction (\ref{2to4_reaction_rhorho})
through $f_{0}$ exchange in the $s$-channel 
as shown in Fig.~\ref{fig:4pi_diagram2} (right diagram).
In the high-energy approximation we can write the amplitude as
shown in (\ref{amplitude_rhorho}) with
%
\begin{equation}
\begin{split}
& {\cal M}^{(f_{0} \to \rho\rho)\,\rho_{3} \rho_{4}}_{\lambda_{a}\lambda_{b}\to\lambda_{1}\lambda_{2}\rho\rho}
\simeq 3 \beta_{\Pom NN}  \, 2(p_1 + p_a)_{\mu_{1}} (p_1 + p_a)_{\nu_{1}}\, 
\delta_{\lambda_{1} \lambda_{a}}\, F_1(t_1)  \;
\frac{1}{4 s_{1}} (- i s_{1} \alpha'_{\Pom})^{\alpha_{\Pom}(t_{1})-1} \\ 
& \quad \quad
\times 
\Gamma^{(\Pom \Pom f_{0})\,\mu_{1} \nu_{1}, \mu_{2} \nu_{2}}(q_{1},q_{2})\,
\Delta^{(f_{0})}(p_{34})\,
\Gamma^{(f_{0} \rho \rho)\, \rho_{3} \rho_{4}}(p_{3},p_{4})\\
& \quad \quad\times 
\frac{1}{4 s_{2}} (- i s_{2} \alpha'_{\Pom})^{\alpha_{\Pom}(t_{2})-1}\,
3 \beta_{\Pom NN}  \, 2 (p_2 + p_b)_{\mu_{2}} (p_2 + p_b)_{\nu_{2}}\, 
\delta_{\lambda_{2} \lambda_{b}}\, F_1(t_2) \,,
\end{split}
\label{amplitude_approx_rhorho}
\end{equation}
where we take for the $f_{0} \rho \rho$ vertex 
the following ansatz
%
\begin{equation}
\begin{split}
&\Gamma^{(f_{0} \rho \rho)\,\rho_{3} \rho_{4}}(p_{3},p_{4}) =\\
&g_{f_{0} \rho \rho}' \,\frac{2}{M_{0}^{3}}\,  
\bigg[ p_{3}^{2} p_{4}^{2} g^{\rho_{3} \rho_{4}}
     - p_{4}^{2} p_{3}^{\rho_{3}} p_{3}^{\rho_{4}}
     - p_{3}^{2} p_{4}^{\rho_{3}} p_{4}^{\rho_{4}}
     + \left( p_{3} \cdot p_{4} \right) p_{3}^{\rho_{3}} p_{4}^{\rho_{4}} \bigg]\,
     F'^{(f_{0} \rho \rho)}(p_{3}^{2},p_{4}^{2},p_{34}^{2})\\
&+ g_{f_{0} \rho \rho}'' \,\dfrac{2}{M_{0}}  \,
\bigg[ p_{4}^{\rho_{3}} p_{3}^{\rho_{4}}-(p_{3} \cdot p_{4}) g^{\rho_{3}\rho_{4}} \bigg]\,
F''^{(f_{0} \rho \rho)}(p_{3}^{2},p_{4}^{2},p_{34}^{2})
\end{split}  
\label{vertex_f0rhorho}
\end{equation}
with $g_{f_{0} \rho \rho}'$ and $g_{f_{0} \rho \rho}''$ being free parameters. 
Different form factors $F'$ and $F''$ are allowed a priori.
The vertex in Eq.~(\ref{vertex_f0rhorho}) fulfils the following relations:
\begin{equation}
\begin{split}
p_{3\, \rho_{3}} \Gamma^{(f_{0} \rho \rho) \rho_{3} \rho_{4}}(p_{3},p_{4}) =0\,, \qquad
p_{4\, \rho_{4}} \Gamma^{(f_{0} \rho \rho) \rho_{3} \rho_{4}}(p_{3},p_{4}) =0\,.
\end{split}
\label{vertex_f0rhorho_aux}
\end{equation}
%

Once evidence for one or more of the $f_{0}$ resonances discussed here
is obtained from RHIC and/or LHC experiments the formulae given
in this appendix should be useful.
Then, it will hopefully be possible to determine empirically the coupling
parameters of the relevant vertices: $\Pom \Pom f_{0}$,
$f_{0} \sigma \sigma$ and $f_{0} \rho \rho$.

\acknowledgments
We are indebted to Leszek Adamczyk,
Lidia G{\"o}rlich, Rados{\l}aw Kycia, Wolfgang Sch{\"a}fer and Jacek
Turnau for useful discussions.
This research was partially supported by 
the MNiSW Grant No. IP2014~025173 (Iuventus Plus),
the Polish National Science Centre Grant 
No. DEC-2014/15/B/ST2/02528 (OPUS)
and by the Centre for Innovation and Transfer of Natural Sciences 
and Engineering Knowledge in Rzesz{\'o}w.

\bibliography{refs}

\end{document}